\documentclass[%
reprint,
onecolumn,
superscriptaddress,
showpacs,
 amsmath,amssymb,
 longbibliography,
 aps,
]{revtex4-1}

\usepackage{graphicx}
\usepackage{dcolumn}
\usepackage{bm}
\usepackage{siunitx,color}
\usepackage{scalerel}
\usepackage{xcolor,color}
\renewcommand{\vec}[1]{\bm{#1}}
\newcommand{\norm}[1]{\left\lVert #1 \right\rVert}
\DeclareMathOperator*{\argmin}{argmin}
\DeclareRobustCommand\smallblacksquare{\scaleobj{0.6}{\blacksquare}}

\def\mysingleq#1{`#1'}

\begin{document}

\preprint{APS/123-QED}

\title[]{Linear genetic programming control for strongly nonlinear dynamics \\ with frequency crosstalk}

\author{Ruiying Li}
\email{ruiying.li@ensma.fr}
\affiliation{Institut Pprime, CNRS -- Universit\'e de Poitiers -- ISAE-ENSMA, Futuroscope Chasseneuil, France\\
}%

\author{Bernd R. Noack}%
\affiliation{Institut Pprime, CNRS -- Universit\'e de Poitiers -- ISAE-ENSMA, Futuroscope Chasseneuil, France\\
}
\affiliation{LIMSI-CNRS, UPR 3251, 91405 Orsay cedex, France\\}
\affiliation{Technische Universit\"{a}t Braunschweig, Braunschweig, Germany\\}
\affiliation{Technische Universit\"{a}t Berlin, Berlin, Germany\\}

\author{Laurent Cordier}
\affiliation{Institut Pprime, CNRS -- Universit\'e de Poitiers -- ISAE-ENSMA, Futuroscope Chasseneuil, France\\
}%

\author{Jacques Bor\'{e}e}
\affiliation{Institut Pprime, CNRS -- Universit\'e de Poitiers -- ISAE-ENSMA, Futuroscope Chasseneuil, France\\
}%

\author{Eurika Kaiser}
\affiliation{University of Washington, Mechanical Engineering Department, \\Seattle, WA 98195, USA}

\author{Fabien Harambat}
\affiliation{PSA Peugeot-Citro\"{e}n, Centre Technique de V\'{e}lizy\\
	V\'{e}lizy-Villacoublay, 78943, France\\}


\begin{abstract}
\label{ToC:Abstract}
We advance Machine Learning Control (MLC), a recently proposed model-free control framework
which explores and exploits strongly nonlinear dynamics in an unsupervised manner.
The assumed plant has multiple actuators and sensors and
its performance is measured by a cost functional.
The control problem is to find a control logic  which optimizes the given cost function.
The corresponding regression problem for the control law is solved by employing \emph{linear genetic programming} as an easy and simple regression solver 
 in a high-dimensional control search space.
This search space comprises open-loop actuation, sensor-based feedback and combinations thereof
--- thus generalizing former MLC  studies \cite{Gautier2015jfm,Parezanovic2016jfm}. 
This methodology is denoted as linear genetic programming control (LGPC).
Focus of this study is the frequency crosstalk between unforced unstable oscillation and the actuation at different frequencies. 
LGPC is first applied to the stabilization of a forced nonlinearly coupled three-oscillator model comprising open- and closed-loop frequency crosstalk mechanisms.
LGPC performance is then demonstrated in a turbulence control experiment,
achieving 22\%  drag reduction for a simplified car model.
For both cases, LGPC identifies the best nonlinear control achieving the optimal performance by exploiting frequency crosstalk.
Our control strategy is suited to complex control problems with multiple actuators and sensors featuring nonlinear actuation dynamics. 
\end{abstract}

\pacs{Valid PACS appear here}
\maketitle

\section{Introduction}
\label{ToC:Intro}
Turbulent flow is characterized by broadband dynamics varying 
from dominant frequencies corresponding to large-scale coherent structures 
to high frequencies corresponding to Kolmogorov microscales.
In a \emph{direct} energy cascade, 
the energy-containing coherent structures transfer the energy 
to small-scale eddies which are dissipated by viscosity.
Inversely, a clustering of coherent structures may yield larger scales at lower frequencies (inverse energy cascade).
Both mechanisms rely on the nonlinearity of Navier-Stokes equations.
This frequency interaction, also called \emph{frequency crosstalk}, 
provides an important challenge and opportunity for flow control: 
the actuation frequency may change the whole spectrum of frequencies 
and thus ultimately affects the mean flow.

The key role of frequency crosstalk in flow control has been reported in numerous studies.
High-frequency forcing using pulsed or synthetic jets or fluidic oscillators \cite{Glezer2002arfm,Cattafesta2011arfm}
has been demonstrated to be able to stabilize the turbulent wakes 
of a circular cylinder \cite{Glezer2005aiaaj}, 
a car model \cite{Barros2016jfm},
a rectangular bluff body \cite{schmidt2015ef} and
an axisymmetric body \cite{oxlade2015jfm}.
It has also been applied on
a flow over a backward-facing step \cite{Vukasinovic2010jfm}, 
a mixing layer \cite{Parezanovic2016jfm} 
and a jet \cite{Samimy2007jfm}.
Low-frequency forcing, on the other hand, 
can either enhance the flow instability manifested 
by the amplified oscillation of vortex shedding \cite{Glezer2005aiaaj, barros2016resonances} 
or attenuate the instability by destroying the formation of shedding \cite{Pastoor2008jfm}.
This frequency-crosstalk effect of actuation 
relies on the nonlinear interactions of high-frequency, low-frequency and the dominant modes of the flow.

Most of the studies mentioned above utilize periodic forcing as control strategy.
Feedback control may increase 
actuation energy efficiency 
by adapting periodic forcing to slowly changing flow conditions \cite{Becker2007aiaaj}.
Feedback may also react on the faster coherent structure dynamics \cite{Brunton2015amr}.
In this case, a physics-based model-based control logic is desirable,
distilling the physical mechanism and its relation to control.
In many cases, this implies that frequency crosstalk is incorporated
in the model which constitutes a big challenge.
Simple examples of such control-oriented models 
describe an actuation at higher or lower frequency
for stabilizing the dominant vortex shedding oscillation \cite{Luchtenburg2009jfm,Sipp2012jfm}.
In general, incorporating multiple frequency crosstalks
in a model-based control strategy constitutes a significant challenge,
both, from a robust modelling and from a control design perspective,
due to the difficulties in the mathematical modelling of the nonlinearities and limited knowledge of flow.
Nevertheless, model-based feedback control has enjoyed many success stories 
for laminar and transitional flows 
for which a linear control theory can be applied \cite{Rowley2006arfm,Bagheri2009amr,Sipp2010amr,Theofilis2011arfm}.
Weakly nonlinear dynamics due to base-flow deformations 
are also easily incorporated in this strategy \cite{gerhard2003aiaa,Fabbiane2016jfm,Thirunavukkarasu2012AIAAj}.

In this study, we target a generic model-free control strategy 
for dynamics with strong nonlinearities 
--- circumventing the challenge to construct corresponding reduced-order models 
and to derive nonlinear control laws. 
Instead, control laws are optimized in the plant with an evolutionary algorithm.
Optimal parameters of open-loop control laws may be determined with a genetic algorithm
 \cite{Benard2016ef}.
The considered search space includes all nonlinear feedback laws 
which are approximated by a finite number of mathematical operations.
Departure point is \emph{Genetic Programming Control (GPC)} \cite{Duriez2016book}.
The determination of feedback control laws is formulated as a regression problem 
in which the controller is optimized with respect to a given cost function.
Genetic programming \cite{koza1992book} is used as a powerful regression technique to explore 
and evolve effective control laws by learning from the training data of experiments or simulations.
Successful applications of GPC include separation control \cite{Gautier2015jfm,Debien2016ef} 
and mixing layer control \cite{Parezanovic2016jfm}.
The innovations in this work include: 
(1) the use of \emph{linear} genetic programming as a simpler algorithm and 
(2) a very general ansatz for control laws incorporating open-loop and sensor-based feedback control.

The paper is organized as follows. 
In Section~\ref{ToC:LGPC}, we present the proposed method and its implementation. 
Then,  in Section \ref{ToC:LGPC results}, we demonstrate LGPC (linear genetic programming control)
to the stabilization of a forced nonlinearly coupled three-oscillator model 
comprising open- and closed-loop frequency crosstalk mechanisms.
In Section \ref{ToC:drag}, LGPC is applied to a turbulence control experiment,
achieving 22\%  drag reduction for a simplified car model.
A landscape of the discovered control laws is visualized 
in Section \ref{ToC:Visualization} to examine its search space topology.
Section \ref{ToC:Conclusion} concludes with a summary and outlook.

\section{Linear genetic programming control}
\label{ToC:LGPC}
We consider a multiple-input multiple-output (MIMO) system  
with the state $\vec{a}\in\mathbb{R}^{N_a}$, 
an input vector $\vec{b}\in\mathbb{R}^{N_b}$ commanding actuation and 
an output vector $\vec{s}\in\mathbb{R}^{N_s}$ sensing the state.
Here, $N_a$, $N_b$ and $N_s$ denote the dimension of the state, 
the number of actuators and sensors, respectively.
The general form of the system reads 
\begin{subequations}
\begin{eqnarray}
\frac{d\vec{a}}{dt}&=\textbf{F}(\vec{a},\vec{b})\\
\vec{s}&=\textbf{G}(\vec{a})\\
\vec{b}&=\textbf{K}(\vec{s}).
\end{eqnarray}
\end{subequations}
The control $\vec{b}$ directly affects the state $\vec{a}$ through a general nonlinear propagator \textbf{F}. 
\textbf{G} is a measurement function comprising the sensor signals $\vec{s}$ as function of the state $\vec{a}$.
The control objective is to construct a MIMO controller $\vec{b}=\textbf{K}(\vec{s})$ 
so that the system has a desirable behaviour. 
Most control objectives can be formulated in a cost function $J(\vec{a},\vec{b})$.
The definition of $J$ depends on the control goal.
For instance, in a drag reduction problem, we define $J$ as the drag power penalized by the actuation power.

Following \cite{Duriez2016book},
the control design is formulated as a regression problem:
find the control law $\bm{b}=\bm{K} ( \bm{s})$
which optimizes a given cost function $J$.
The  cost only depends on the control law, or, symbolically
 $J\left(\textbf{K} (\vec{s}) \right) $
for a well-defined initial value problem
or statistically stationary actuation response.
Summarizing, the control task is transformed to an optimization problem 
via cost minimization and is equivalent to finding $\textbf{K}^{\text{opt}}$ such that
\begin{equation}
\textbf{K}^{\text{opt}}(\vec{s})=\mathop{\argmin}_{\textbf{K}}{J(\textbf{K} (\vec{s}))}.
\end{equation} 

The sensor-feedback law maps $N_s$ sensor signals 
onto $N_b$ actuation commands.
Such feedback can be expected 
to be approximated by a finite number of elementary operations
($+,-,\times,\div,...$) acting on the sensor signals $\bm{s}$ 
and finite number of fixed constants.
Thus, the search space of permissible control laws is finite,
yet of astronomical cardinality. 
Hence, an exhausting testing in an experiment or numerical calculation is not an option.
Instead, we employ genetic programming (GP) as powerful evolutionary search algorithm.
GP yields optimal or near-optimal control laws in the search space 
with high probability for suitable parameters,
yet with no mathematically assured convergence.
The original tree-based genetic programming (TGP) 
formulates the mapping by a binary tree structure \cite{koza1992book}.
Here, we propose to apply a more recent alternative to TGP, 
called linear genetic programming (LGP) \cite{brameier2007linear}.
TGP and LGP are equivalent in the sense that any LGP-law can be expressed in TGP and vice versa. 
The difference is the linear versus recursive coding of LGP and TGP, respectively. 
LGP is much easier to code and implement 
in systems with multiple actuators and multiple sensors.
As presented before, we refer to this method as \emph{linear genetic programming control} (LGPC). 
For details of LGPC, see \cite{Li2016ef2}.

\begin{figure}
	\centering	\includegraphics[scale = 0.7]{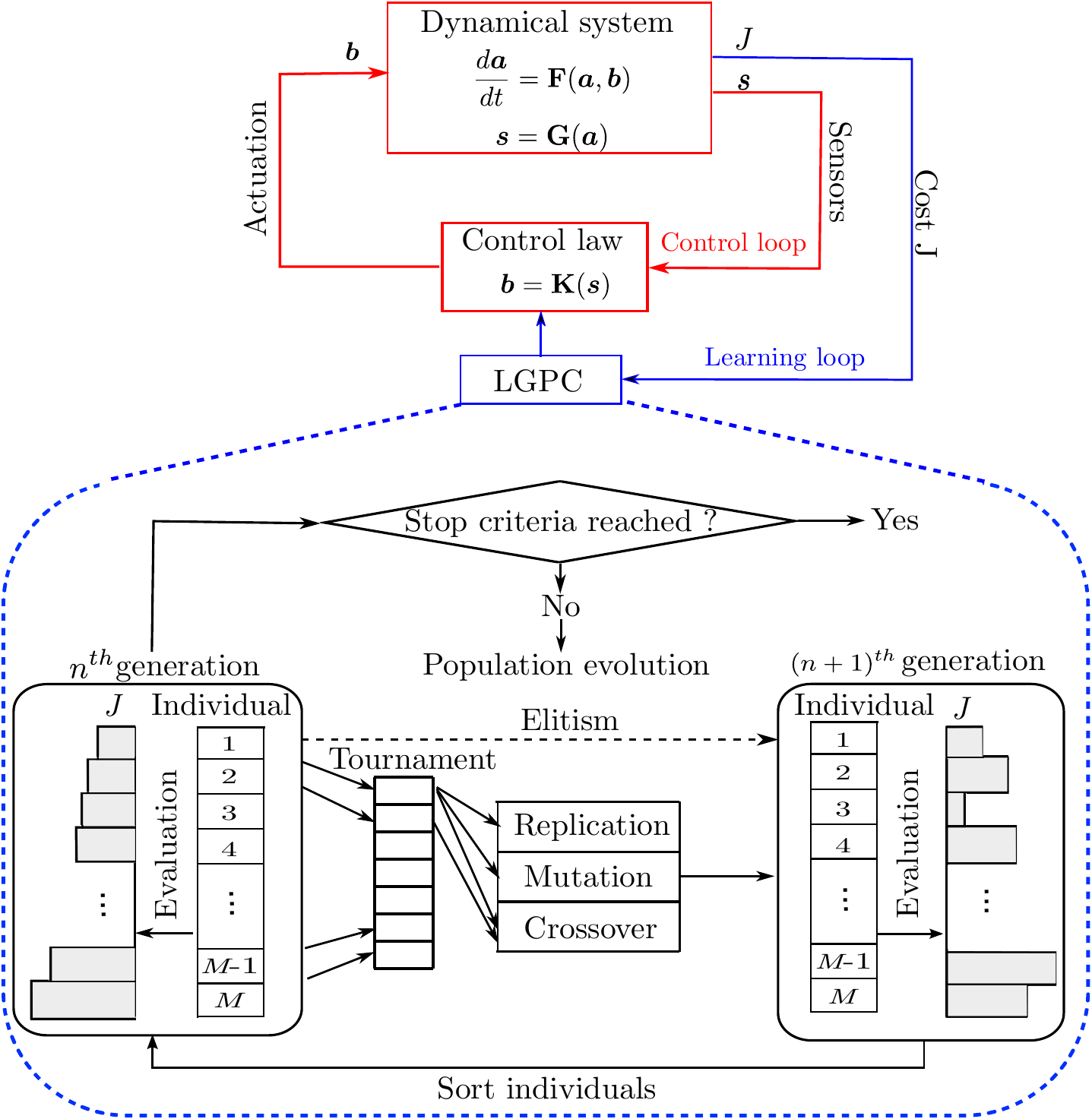}
	\caption{LGPC implementation. 
                The real-time closed-loop control is  performed in the inner  loop (red lines).
		The control plant feeds back the sensor output $\vec{s}$ to the control law. 
		This control law proposed by LGPC computes the actuation command 
                based on $\vec{s}$ and sends it back to the plant. 
		A cost $J$ is attributed to the control law after its evaluation during the time $T$. 
		In the outer learning loop, LGPC  uses these costs $J$ to evolve the new population of control laws. 
         The LGPC learning process is depicted in the lower part. 
                On the leftmost side, an evaluated generation with $M$ individuals is sorted 
		in ascending order based on $J$. 
		If the stopping criterion is met, the learning process is terminated.
		If not, the next generation (on the rightmost side) 
                is evolved by genetic operators (elitism, replication, mutation, and crossover). 
		After being evaluated, this generation is sorted as indicated by the arrow at the bottom. 
		We repeat the process from left to right until the stopping criterion is met.}
	\label{fig:Algorithm}
\end{figure}

The implementation of LGPC for feedback control 
is sketched in Fig.~\ref{fig:Algorithm}. 
The fast real-time control occurs in the inner loop 
with a control law proposed by LGPC. 
The control law is evaluated in the dynamical system over an evaluation time $T$. 
Then, a cost $J$ is measured quantifying the performance of the control law. 
The cost value for each control law is sent to the slow outer learning loop, 
where LGPC evolves these laws.

\graphicspath{{./Figures/}}

The learning process of LGPC is detailed in the lower part of Fig.~\ref{fig:Algorithm}. 
An initial population of control law candidates, called \emph{individuals,} 
is generated randomly like in a Monte-Carlo method (%
see Sec.~3.3
in \cite{Li2016ef2}). 
Each individual is evaluated in the inner loop and a cost $J$ is attributed to them. 
After the whole generation is evaluated, 
its individuals are sorted in ascending order based on $J$. 
The next generation of individuals is then evolved 
from the previously evaluated one by genetic operators (elitism, replication, mutation, and crossover).
Elitism is a deterministic process 
which copies a given number of top-ranking individuals 
directly to the next generation. 
This ensures that the next generation will not perform worse than the previous one. 
The remaining genetic operations are stochastic in nature 
and have specified selection probabilities. 
The individual(s) used in these genetic operators is (are) selected by a tournament process: $N_t$ randomly chosen individuals compete in a tournament and the winner(s) (based on $J$) is (are) selected.
Replication copies a statistically selected number of individuals to the next generation.
Thus better performing individuals are memorized.
Crossover involves two statistically selected individuals 
and generates a new pair of individuals by exchanging randomly their instructions. 
This operation contributes to breeding better individuals 
by searching the space around well-performing individuals. 
In the mutation operation, random elements  
in the instructions of a statistically selected individual are modified. 
Mutation serves to explore potentially new and better minima of $J$. 
After the new generation is filled, the evaluation of this generation can be pursued in the plant. 
This learning process will continue until some stopping criterion is met. 
Different criteria are used.
Ideally, the process is stopped when a known global minimum is obtained (which is unlikely in an experiment).
Alternatively, the evolution terminates upon too slow improvement from one generation to the next
or when a predefined maximum number of generations is reached. 
By definition, the targeted optimal control law is the best individual of the last generation.

LGPC can also be used to explore open-loop control 
by including time-periodic functions $\vec{h}$ 
in the inputs of control law, i.e. $b=\textbf{K}(\vec{h})$. 
This method permits to search a much more general multi-frequency control 
which is hardly accessible to a parametric study of single frequency.
Furthermore, the range of LGPC can be extended by comprising both the sensors $\vec{s}$ 
and time-periodic functions $\vec{h}$ into the inputs of $\vec{K}$. 
This results in a non-autonomous control law $\vec{b}=\vec{\textbf{K}}(\vec{s},\vec{h})$. 
This generalization permits to select between open-loop actuation $\vec{b}=\vec{\textbf{K}}(\vec{h})$, 
sensor-based feedback $\vec{b}=\vec{\textbf{K}}(\vec{s})$ or combinations thereof $\vec{b}=\vec{\textbf{K}}(\vec{s},\vec{h})$ 
depending on which performs better. 
In the following, we term the approach optimizing open-loop frequency combinations $b=\textbf{K}(\vec{h})$ as LGPC-1.
The approach to optimize autonomous controllers $\vec{b}=\vec{\textbf{K}}(\vec{s})$ is referred to as LGPC-2.
The generalized non-autonomous control design $\vec{b}=\vec{\textbf{K}}(\vec{s},\vec{h})$ is denoted as LGPC-3.


\section{Model of three coupled oscillators}
\label{ToC:LGPC results}
\begin{figure}
	\centering
	\includegraphics[scale = 0.8]{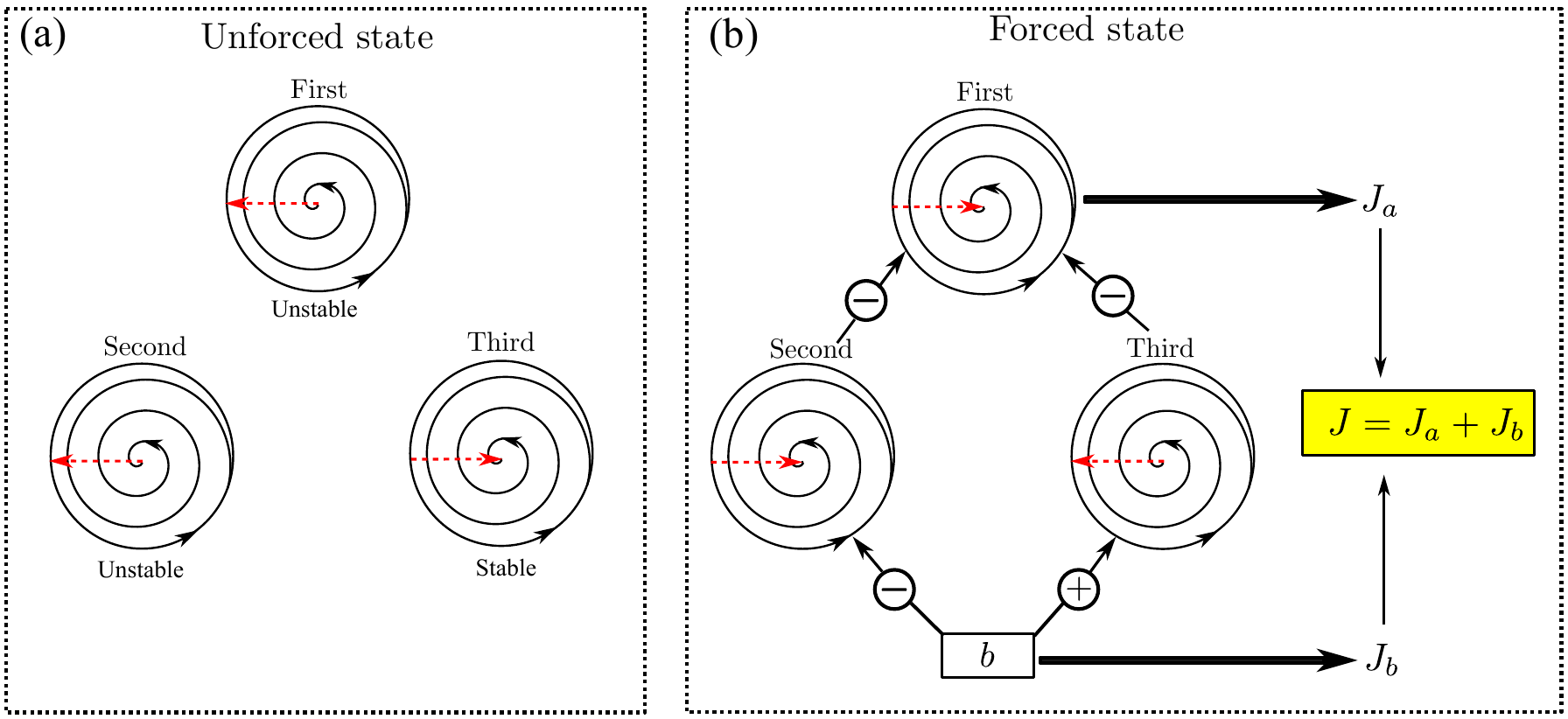}
	\caption{Illustration of the three-oscillator model: (a) unforced state and (b) forced state.
	The red dashed arrows indicate the tendency of amplitudes of oscillators.
	The sign `$-$' and `$+$' in (b) represent the suppression and excitation of oscillators, respectively.}
	\label{fig:Osc_sketch}
\end{figure}

In this section, 
we illustrate LGPC to stabilize a forced dynamical system 
with three nonlinearly coupled oscillators at three incommensurable frequencies 
extending the generalized mean-field model \cite{Luchtenburg2009jfm}
(see Chapter 5 of \cite{Duriez2016book}).
The goal is to stabilize the first unstable, amplitude-limited oscillator,
while the forcing is performed on the second and third oscillator 
(see Fig. \ref{fig:Osc_sketch}).
The second oscillator has also unstable, amplitude-limited dynamics
and destabilizes the first oscillator. 
The third oscillator has linear stable dynamics and has a stabilizing effect on the first.
The stabilization of the first oscillator can be performed 
by closed-loop suppression of the second oscillator 
or open-loop excitation of the third one.
In the following, 
we formulate the control problem mathematically (Section \ref{ToC:Model:Problem}), 
parametrically explore the effect of periodic forcing
like in many turbulence control experiments (Section \ref{ToC:Model:Periodic}), 
and apply LGPC (Section \ref{ToC:Model:LGPC}).
\graphicspath{{./Figures/}}

\subsection{Problem formulation}
\label{ToC:Model:Problem}
The system has three oscillators at frequency $\omega_1=1$, $\omega_2=\pi$ and $\omega_3=\pi^2$, 
the coordinates of which being $(a_1,a_2)$, $(a_3,a_4)$ and $(a_5,a_6)$, respectively. 
The evolution equation of the state $\vec{a}=(a_1, a_2, \ldots, a_6)$ reads:
\begin{equation}
\begin{aligned}
\dfrac{da_1}{dt}&=\sigma_1 a_1-a_2 &\hspace{20pt}\dfrac{da_3}{dt}&=\sigma_2 a_3-\pi a_4  &\hspace{20pt}\dfrac{da_5}{dt}&=\sigma_3 a_5-\pi^2 a_6\\
\dfrac{da_2}{dt}&=\sigma_1 a_2+a_1 &\hspace{20pt}\dfrac{da_4}{dt}&=\sigma_2 a_4+\pi a_3+b &\hspace{20pt}\dfrac{da_6}{dt}&=\sigma_3 a_6+\pi^2 a_5+b\\
\sigma_1&=-r_1^2+r_2^2-r_3^2 &\hspace{20pt}\sigma_2&=0.1-r_2^2&\hspace{20pt}\sigma_3&=-0.1,
\end{aligned}
\label{eq:3osc}
\end{equation}
where $r_1^2=a_1^2+a_2^2$, $r_2^2=a_3^2+a_4^2$ and $r_3^2=a_5^2+a_6^2$ 
denote the fluctuation level of the three oscillators, respectively. 
The growth rate for each oscillator is denoted by $\sigma_i, i=1,\ldots,3$.
Without forcing $b \equiv 0$, 
the first and second system are linearly unstable 
and damped by a Landau-type cubic term to asymptotic amplitudes $r_1^u=r_2^u=\sqrt{0.1}$.
Here, and in the following, the superscript \mysingleq{$u$} refers to asymptotic values for unforced dynamics.  
The third system is linear and stable, i.e.\ converges to the vanishing amplitude $r_3^u=0$.
The forcing $b$ is only applied on the second and third oscillators.
A linearization of Eqs.~\eqref{eq:3osc} around the fixed point $\vec{a}=\vec{0}$
yields a system in which the first oscillator is uncontrollable.

The effect of the forcing on the first oscillator 
can be inferred from the growth rate formula for $\sigma_1$ (see first column in Eqs. \eqref{eq:3osc}). 
The fluctuation level $r_2$ of the second system destabilizes the first oscillator,
while the third system stabilizes it with increasing fluctuation level $r_3$.
Hence, stabilization of the first oscillator may be achieved 
by exploiting one of two frequency crosstalk mechanisms:
stabilizing the second system or exciting the third one.
Evidently stabilization of the second system requires feedback $b=K(\vec{a})$
while excitation of the stable oscillator 
can be performed with periodic forcing $b(t)=B \sin\left(\pi^2 t\right)$
at the resonance frequency and sufficiently large amplitude $B$.

The cost function to be minimized is the averaged energy of 
the unstable oscillator $J_a=\overline{a_1^2+a_2^2}$ 
penalized by the actuation cost $J_b=\overline{b^2}$.
Here, the temporal averaging is indicated by the overbar.
Without forcing, $J_a^u=\left ( r_1^u \right)^2$ and $J_b\equiv 0$.
We normalize the total cost by the unforced value $J_a^{u}$ of the first oscillator 
to characterize the relative benefit of actuation:
\begin{equation}
J=\dfrac{J_a+\gamma J_b}{J_a^{u}},
\label{eq:J}
\end{equation}
with $\gamma=1$ as the penalization coefficient.
By definition, $J=1$ for the unforced system.

The numerical evaluation of $J$ is based on the integration of the dynamical system \eqref{eq:3osc}
with the initial condition $\vec{a}(0)=(0.1, 0, 0.1, 0, 0.1, 0)$ at $t=0$.
In the first 10 periods of the target oscillator, 
\textit{i.e.}\ for $t\in[0,t_0]$ with $t_0=10 \frac{2\pi}{\omega_1}=20 \pi$, no forcing is applied
and the system converges to unforced quasi-periodic dynamics $(r_1^u)^2=0.1$, $(r_2^u)^2=0.1$, $r_3^u=0$.
The cost functional is evaluated in the next 500 periods, $t\in[20\pi, 1020\pi]$.
This time interval contains an actuated transient 
but is dominated by the post-transient dynamics,
\textit{i.e.}\  sufficient for statistical averaging.

\subsection{Open-loop periodic forcing}
\label{ToC:Model:Periodic}
\begin{figure}[htb]
	\centering
	\includegraphics[scale = 0.48]{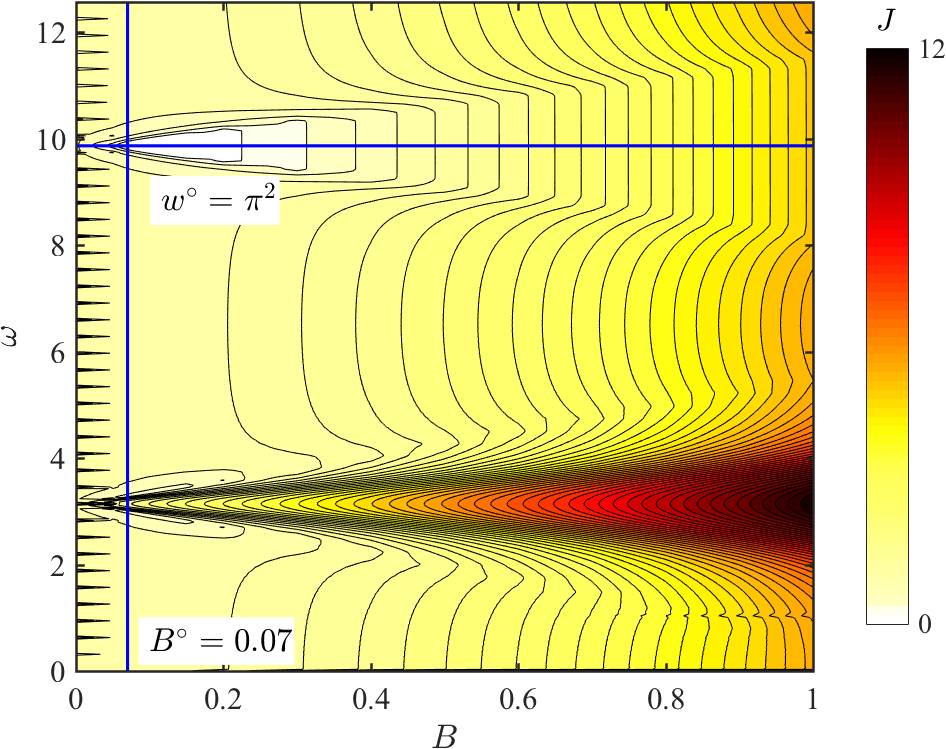}
	\caption{Colormap of cost value $J$ under the periodic forcing $b(t)=B\sin(\omega t)$. }
	\label{fig:scanofw}
\end{figure}

First, open-loop periodic forcing is studied,
following a practice of many turbulence control experiments.
The goal is to minimize the cost function Eq. \eqref{eq:J} 
with periodic forcing $b_{\text{OL}}(t)=B\sin(\omega t)$
employing a parametric variation of the amplitude $B$ and frequency $\omega$ 
in the range of $[0,1]$ and $[0,4\pi]$. respectively.
The performance (Eq. \eqref{eq:J}) at amplitude $B$ and frequency $\omega$
is scanned with increments $0.01$ and $0.01 \pi$, respectively.
The corresponding colormap of $J$ is shown in Fig.~\ref{fig:scanofw}.
This figure displays a local minimum of $J^{\circ}=0.031$.
The corresponding parameters are denoted by the superscript `$\circ$' in the following.
The low value indicates a stabilization by over one order of magnitude 
in the fluctuation level, accounting for the actuation expense.
The minimum $J$ is reached at the eigenfrequency 
of the third oscillator $\omega^{\circ}=\pi^2$,
as $\sigma_1<0$ for $r_3^2>0.1$, 
numerically observing that the second oscillator is hardly affected by the forcing at a non-resonant frequency,
$r_2^{\circ} \approx r_2^u =\sqrt{0.1}$.
The optimal amplitude $B^{\circ}=0.07$ is numerically determined as the best trade-off 
between the achieved stabilization and actuation cost.
This amplitude leads to $r_3^2 \approx 0.12$ and $\sigma_1 \approx -0.02$.
For a larger time evaluation horizon, 
the current results suggest a better performance at lower actuation $B\approx 0.05$
leading to $r_3^2 \approx 0.1$ 
which just neutrally stabilizes the first oscillator $\sigma_1 \approx 0$, 
exploiting that the second oscillator is unaffected by forcing.
The corresponding analytical approximations are described in Chapter 5 of \cite{Duriez2016book}.
\graphicspath{{./Figures/}}

On the other hand, the maximal $J$ value is associated with the forcing 
at the eigenfrequency of the second oscillator $\omega=\pi$, 
as the excitation of $r_2$ leads to $\sigma_1>0$, resulting in an increase of $r_1$.
These results show that the enabler of open-loop control is the third oscillator rather than the second.

\begin{figure}[htb]
	\centering
	\includegraphics[scale = 0.7]{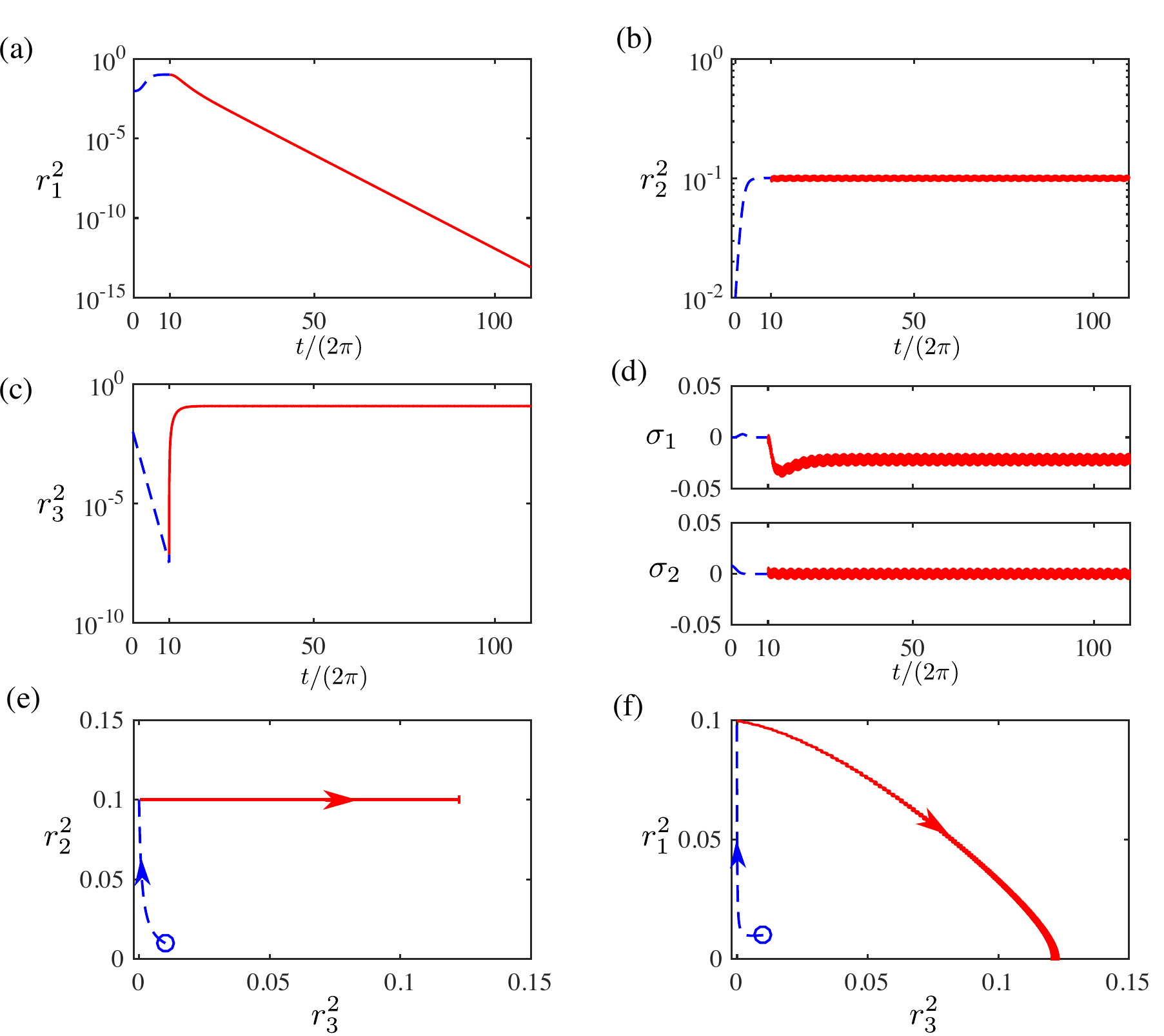}
	\caption{Dynamics of the model system \eqref{eq:3osc} with the optimal periodic forcing $b^{\circ}(t)=0.07\sin(\pi^2t)$ applied at $t/(2\pi)\geqslant10$. Unforced state: blue dashed line; forced state: red line. 
		(a-d) Time evolution of $r_1^2$, $r_2^2$, $r_3^2$, $\sigma_1$ and $\sigma_2$. Only the first 110 periods are shown here for clarity.
		(e) Phase portrait of $r_2^2$ against $r_3^2$ and (f) $r_1^2$ against $r_3^2$.
		}
	\label{fig:best ol}
\end{figure}

The unforced transient and actuated dynamics of the system 
are illustrated in Fig.~\ref{fig:best ol} 
under the optimal periodic forcing $b^{\circ}(t)=0.07\sin(\pi^2 t)$. 
The unforced state during the time window $t\in[0,20\pi]$ is depicted 
by a blue dashed line and the forced one at $t>20\pi$ by a red curve.
For clarity, only the first 110 periods are shown in Fig.~\ref{fig:best ol} (a-d).
Fig.~\ref{fig:best ol} (e,f) covers the whole time interval $t\in[0, 1020\pi]$.
When unforced, the unstable oscillators self-amplify 
towards the limit cycle $r_1^2=r_2^2=0.1$, 
whilst the stable oscillator vanishes to $r_3^2=0$.
Convergence is implied by   $\sigma_1=0$ and $\sigma_2=0$.
Once $b$ starts at $t_0=20\pi$, 
$r_3$ is rapidly excited to an energy level of $r_3^2=0.12$, 
while $r_2$ keeps its original fluctuation level $r_2^2=0.1$.
The resulting system yields $\sigma_1<0$ 
which leads consequently to the stabilization of $(a_1,a_2)$, \emph{i.e.} $r_1^2 \approx 0$. 
The phase portraits in Fig.~\ref{fig:best ol}(e) and (f) 
illustrate the interactions between different oscillators. 
The circle indicates the initial point and the arrows the time direction.
The forced trajectories represent low-pass filtered data,
\textit{i.e.}\ do not resolve cycle-to-cycle variation.
In particular, Fig.~\ref{fig:best ol}(f) shows clearly that $r_1^2$ decreases with the increase of $r_3^2$, 
corroborating that a high-frequency forcing stabilizes a low-frequency unstable oscillator 
via frequency crosstalk.
\graphicspath{{./Figures/}}

\subsection{Results of LGPC}
\label{ToC:Model:LGPC}
LGPC is applied to solve the control problem
of Section \ref{ToC:Model:Problem}.
For all LGPC tests,
up to $N=50$ generations with $M=500$ individuals in each are evaluated.
Hereafter, we denote the cost value of the $m$th individual in the $n$th generation 
by $J_m^n$ $(m=1,\ldots,M; n=1,\ldots,N)$.
After generating the individuals, 
each is pre-evaluated based on the state $\vec{a}$ of the unforced system.
The resulting actuation command is an indicator for their feedback control performance.
If no actuation ($b=0, \forall t$) is obtained in the pre-evaluation, 
this individual cannot change the unforced state.
As a consequence,
the individual 
is not subjected to a testing and is assigned a high cost value.
This pre-evaluation step saves numerical testing time.

The parameters of linear genetic programming are similar to those of most GPC studies 
(see, e.g.\ the textbook \cite{Duriez2016book}). 
Elitism is set to $N_e=1$, \textit{i.e.}\ the best individual of a generation is copied to the next one.  
The probabilities for replication, crossover and mutation are 10\%, 60\% and 30\%, respectively. 
The individuals on which these genetic operations are performed 
are determined from a tournament selection of size $N_t=7$. 
The instruction number in the initial generation 
is selected  between $2$ to $30$ with a uniform probability distribution. 
In the following generations, the maximum instruction number for each individual is capped by $100$.
Elementary operations comprise $+,\, -,\, \times,\, \div,\, \sin,\, \cos,\, \tanh$ and $ \ln$. 
The operation $\div$ and $\ln$ are protected, 
\textit{i.e.}\ the absolute value of the denominator of $\div$ is set to $10^{-2}$ when $|x|<10^{-2}$.
Similarly, $\ln(x)$ is modified to $\ln(|x|)$ where $|x|$ is set to $10^{-2}$ when $|x|<10^{-2}$.
In addition, we choose six random constants  in the range $[-10,10]$ with uniform probability distribution.

In the following, we introduce successively the results of
open-loop multi-frequency forcing LGPC-1 (Section \ref{ToC:3_osc_LGPC_1}),
full-state feedback control LGPC-2 (Section \ref{ToC:3_osc_LGPC_2})
and non-autonomous control LGPC-3 (Section \ref{ToC:3_osc_LGPC_3}).

\subsubsection{LGPC-1}
\label{ToC:3_osc_LGPC_1}

\begin{figure}
	   	\centering
	   	\includegraphics[scale = 0.45]{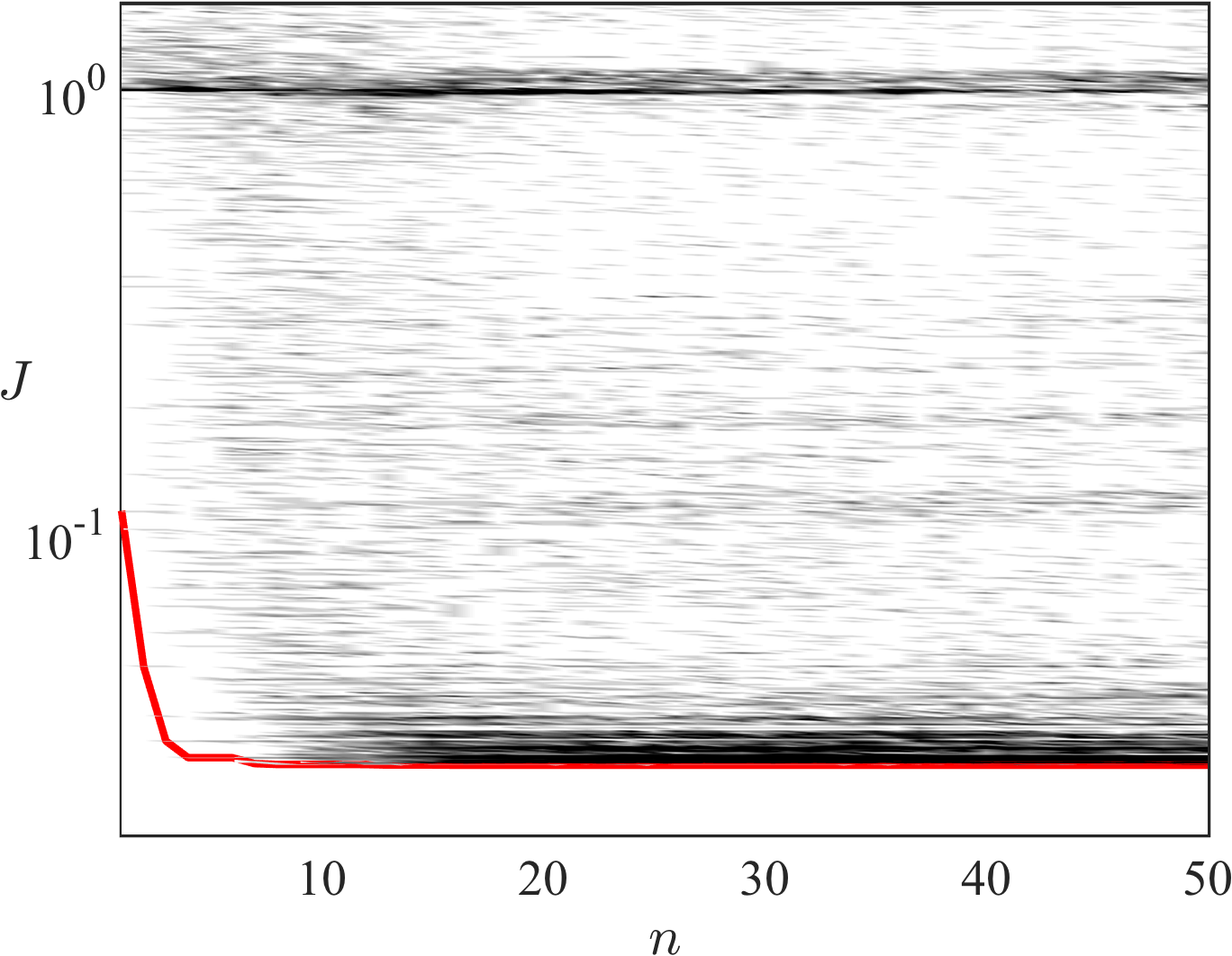}
	   	\caption{‘Spectrogram’ of all computed $J_m^n$ $(m=1,\ldots,M; n=1,\ldots,N)$ for LGPC-1. 
	   	For each generation $n$, $J_m^n$ is ordered with respect to their cost $J_1^n \leqslant J_2^n \leqslant \ldots \leqslant J_M^n$. 
	   	The color shows the distribution of cost values.
	   	Darker color indicates larger proportion.
	   	The red line highlights the best cost value of each generation $J_1^n$. }
	   	\label{fig:Convergence_3osc_LGPC1}
	   \end{figure}

	First we search for generalizing the open-loop control
	by including the best periodic forcing at all eigenfrequencies, 
	i.e. $b=K(\vec{h})$ 
	where $\vec{h}= \left( h_1,h_2,h_3 \right)
	= \left( \sin(t), \sin(\pi t), \sin(\pi^2t) \right)$.
	This approach, called LGPC-1, contains the best periodic forcing frequency $\omega^\circ=\pi^2$, thus it should be at least as good than the optimal periodic forcing $b^\circ$.
	Figure~\ref{fig:Convergence_3osc_LGPC1} displays the `spectrogram' of the cost values for the whole collection of control laws.
	   \graphicspath{{./Figures/}}
	Each generation $n$ is seen to consist of a large range of cost values.
	The decreasing $J$ values towards the right bottom with increasing generation 
	evidences the learning of increasingly better control laws.
	The best cost value of each generation is highlighted by a red line.
	The best individual ($m=1$) in the last generation ($n=50$) reads
	\begin{equation}
	b^{\odot}(t)=-0.37\sin\big(-0.18\sin(\pi^2 t)\big).
	\label{eq:b_lgpc2}
	\end{equation}
	Here, and in the following, the superscript `$\odot$' refers to LGPC-1.
    When applying a first order approximation on $b^{\odot}$, 
    we get $b^{\odot}(t)\approx0.067\sin(\pi^2 t)$.
    This expression resembles that of the optimal periodic forcing $b^{\circ}(t)=0.07\sin(\pi^2 t)$, and leads to a slightly better cost $J^\odot=0.03$ as a better amplitude with a higher precision is explored by LGPC-1. 
   	The dynamics of the system with $b^{\odot}$ are similar to Fig.~\ref{fig:best ol} and are not shown here for brevity.

If we increase the precision of $B$ to 0.001 in the parameter scan of the periodic forcing in Section~\ref{ToC:Model:Periodic},
we should find the same result.
However, the number of evaluations raises to $N_{B}\times N_{\omega}=1001\times401=401000$  ($N_B$ and $N_{\omega}$ being the number of the amplitudes and frequencies to be tested, respectively)
which is 16 times that of LGPC-1 which equals $M\times N=500\times50=25000$.
In summary, LGPC-1 identifies automatically the optimal frequency  $\omega^\odot=\pi^2$ and the optimal amplitude $B^\odot=0.067$
by employing less time than that for the periodic forcing with an exhaustive parameter sweep.

\subsubsection{LGPC-2}
\label{ToC:3_osc_LGPC_2}
\begin{figure}
\centering
\includegraphics[scale = 0.45]{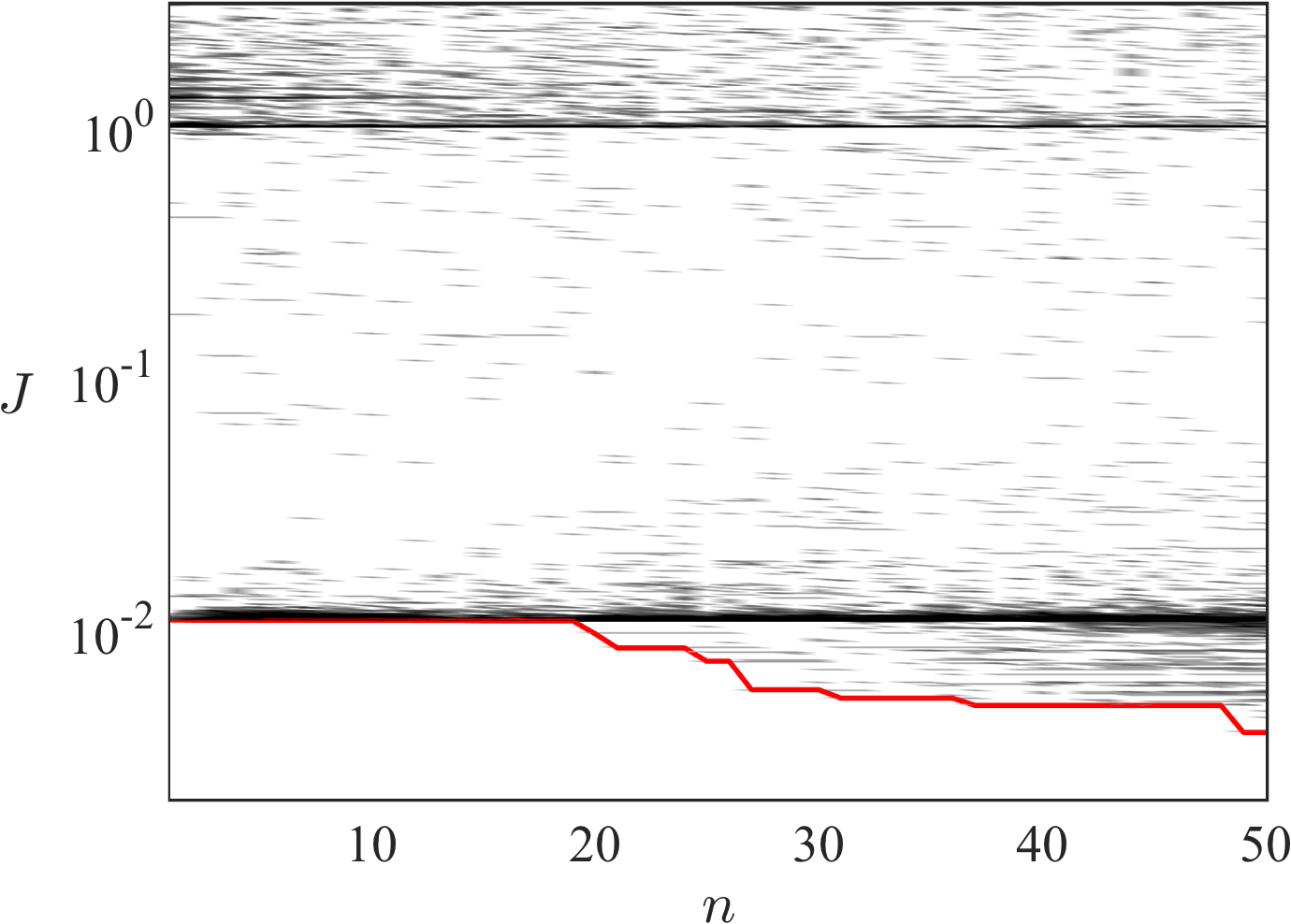}
\caption{Same as Fig.~\ref{fig:Convergence_3osc_LGPC1}, but for LGPC-2.
}
\label{fig:Convergence_3osc_LGPC2}
\end{figure}

\begin{figure}[htb]
	\centering
	\includegraphics[scale = 0.55]{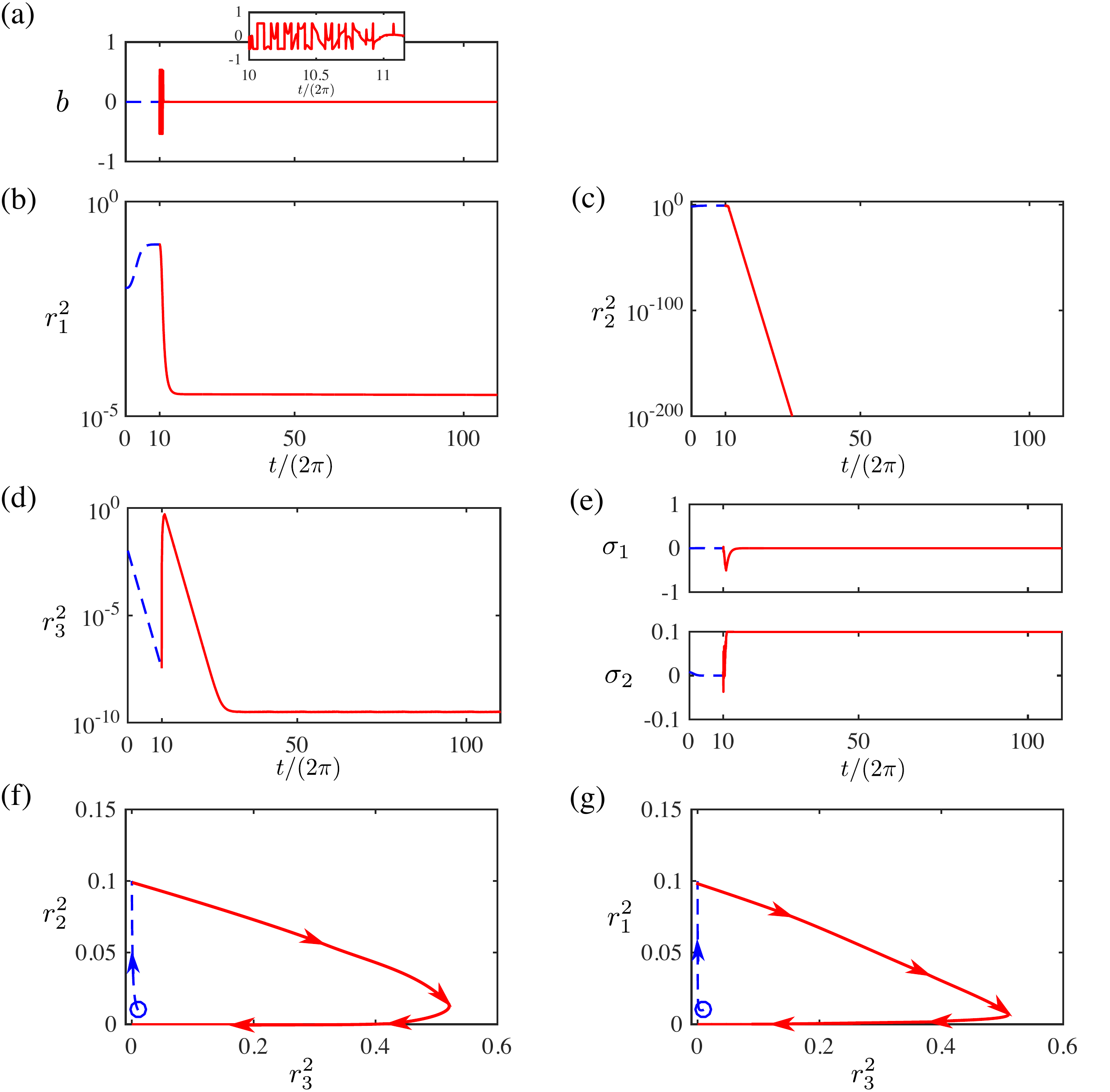}
	\caption{Dynamics of the dynamical system \eqref{eq:3osc} 
		with the LGPC-2 control $b^{\smallblacksquare}$ applied at $t/(2\pi)\geqslant10$. 
		Unforced state: blue dashed line; forced state: red line. 
		(a-e) Time evolution of $b$, $r_1^2$, $r_2^2$, $r_3^2$, $\sigma_1$ and $\sigma_2$. Only the first 110 periods are shown here for clarity.
		(f) Phase portrait of $r_2^2$ against $r_3^2$ and (g) $r_1^2$ against $r_3^2$.}
	\label{fig:best cl}
\end{figure}

Next,  an autonomous full-state feedback law (LGPC-2) is optimized,
$$ b=K(\vec{a})=K(a_1,a_2,a_3,a_4,a_5,a_6). $$
The `spectrogram' of the cost values is shown in Fig.~\ref{fig:Convergence_3osc_LGPC2}.
The successive jumps of the best cost value for each generation (red line) reflect the evolution process to better individuals.
\graphicspath{{./Figures/}}
The targeted LGPC-2 feedback law, \textit{i.e.}\ the best individual 
in the last generation, reads as follows:
\begin{equation}
b^{\smallblacksquare}=\tanh\Bigg(\sin\Bigg(\tanh\Bigg(\tanh\bigg(\tanh\Big(\big(\ln(a_4)+\dfrac{5.8}{\frac{a_6}{1-a_6}}a_4\big)a_4\Big)\bigg)\Bigg)\Bigg)\Bigg).
\label{eq:b_lgpc}
\end{equation}
Here, and in the following, the superscript
`$\smallblacksquare$' refers to LGPC-2.
The corresponding cost $J^{\smallblacksquare}=0.0038$
is more than seven times better than the value achieved 
with optimal open-loop control $b^{\circ}$.
Closed-loop control $b^{\smallblacksquare}$ leads to both,
a smaller fluctuation level $J_a$ and a lower actuation energy $J_b$.
The corresponding dynamics are depicted in Fig.~\ref{fig:best cl}. 
\graphicspath{{./Figures/}}
Instead of the regular excitation of periodic forcing, 
Fig.~\ref{fig:best cl} (a) shows that $b^{\smallblacksquare}$ gives
a strong initial \mysingleq{kick} on the system by exciting the third oscillator 
to a high energy level of $r_3^2=0.5$ (see Fig.~\ref{fig:best cl} (d), (f) and (g)), 
while simultaneously stabilizing the second oscillator, $r_2^2\approx 0$ (see Fig.~\ref{fig:best cl} (c) and (f)).
The first oscillator exhibits consequently a fast decay as $\sigma_1$ has decreased to $\sigma_1=-0.5$ 
due to the change in $r_2^2$ and $r_3^2$ (see Fig.~\ref{fig:best cl} (b), (e) and (g)).
This fast transient takes about one period $\Delta t=2\pi$,
see the close view of forcing $b$ in Fig.~\ref{fig:best cl} (a).
It should be emphasized that LGPC-2 discovers and exploits both frequency crosstalk mechanisms,
the excitation of the third oscillator for a quick transient and the suppression of the second oscillator
to sustain the low fluctuation level of the target dynamics.

Following this fast transient, 
the first and second oscillators enter into a quasi-stable state at nearly vanishing fluctuation levels.
Subsequently, the control command vanishes as full-state feedback shows  no need to actuate after the energy is defeated. 
With vanishing $b$, the third oscillator decays exponentially fast.
This transient process converges to the fixed point
as  depicted  in Fig.~\ref{fig:best cl} (f) and (g).
Now, the first oscillator has a stabilizing growth rate $\sigma_1 \approx -r_1^2$.
LGPC-2 shows an example of feedback control better than the open-loop control.
With only a tiny investment of actuation energy at the very beginning of the control, 
the whole system remains stabilized without actuation even after thousands of periods.

It should be noted that closed-loop control 
is not necessarily better than open-loop actuation.
Suppose the growth-rate of the first oscillator reads
\begin{equation}
\sigma_1 = 0.1 - r_1^2 + r_2^2/100 - r_3^2.
\label{Eq:Example_sigma1}
\end{equation}
In this case, exciting the third oscillator is the only effective stabilizing mechanism
and this excitation can already be done with open-loop forcing.

\subsubsection{LGPC-3}
\label{ToC:3_osc_LGPC_3}
\begin{figure}[htb]
	\centering
	\includegraphics[scale = 0.55]{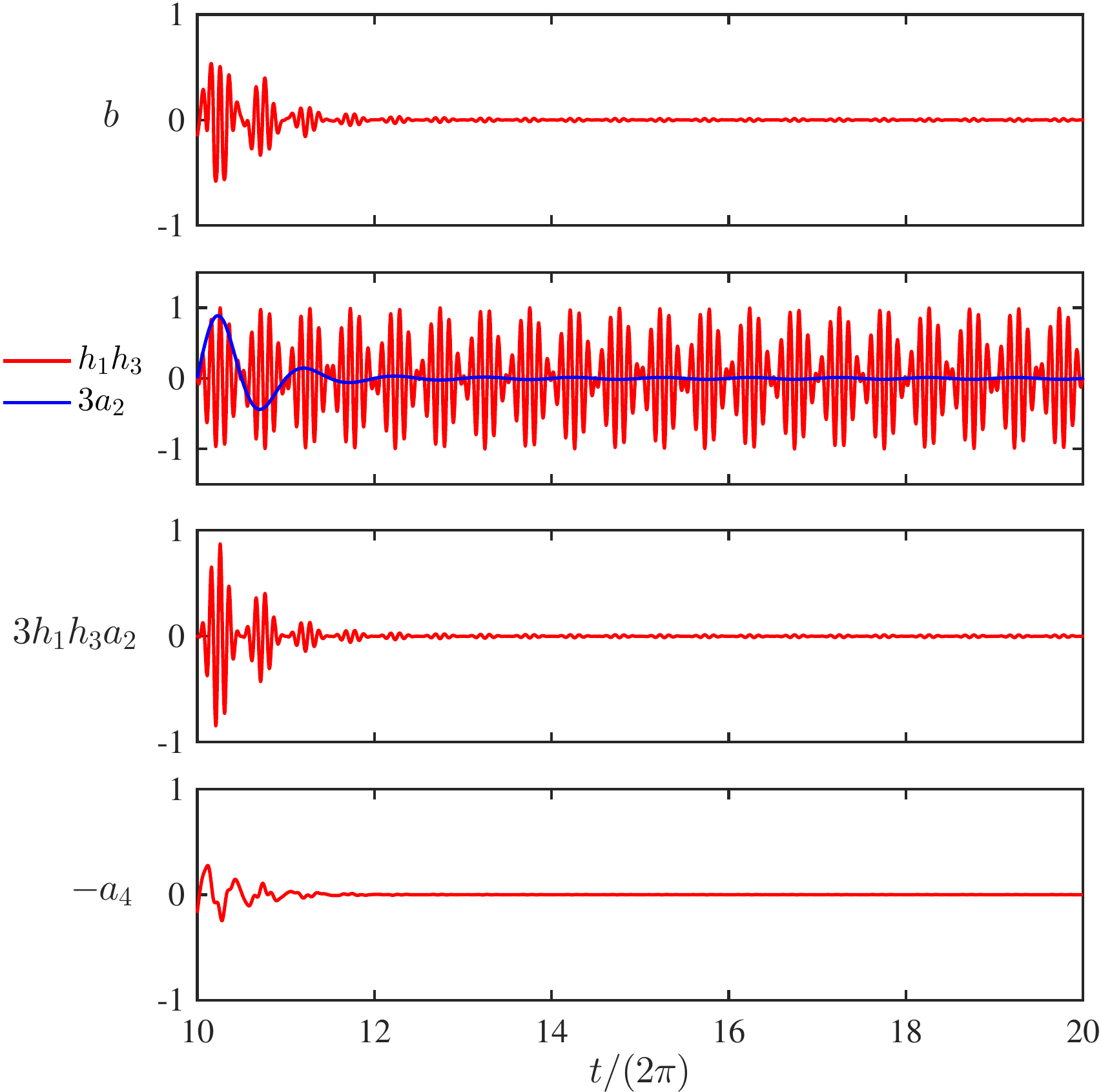}
	\caption{Time evolution of $b^{\bullet}$, $h_1 h_3$, $a_3$ and $a_4$.}
	\label{fig:Viz_b_LGPC3}
\end{figure}

Finally, we explore a more general class of control laws 
which combines full-state feedback $\vec{a}$ and
the best periodic forcing at all eigenfrequencies $\vec{h}= \left( \sin(t), \sin(\pi t), \sin(\pi^2t) \right)$,
as discussed in Section  \ref{ToC:LGPC}.
Then, the generalized LGPC-3 control law $b=K(\vec{a}, \vec{h})$
includes 
the pure full-state feedback 
and the best periodic forcing frequency $\omega^\circ$.
Hence, it should be at least as good than LGPC-2. 
The learning process is similar to Fig.~\ref{fig:Convergence_3osc_LGPC2}, thus we  do not show the convergence of cost values here for brevity.
The optimal control law from LGPC-3 reads
\begin{equation}
b^{\bullet}(t)=\tanh\Bigg(\sin\bigg(\tanh\Big(\big(3 a_2 \sin(t)\sin(\pi^2 t)-a_4\big)\Big)\bigg)\Bigg).
\label{eq:b_lgpc3}
\end{equation}
Here, and in the following, the superscript `$\bullet$' refers to LGPC-3 results.
This control law achieves a better cost value $J^{\bullet}=0.0025$
compared to LGPC-1 with similar dynamics. 
Hence, the results are not detailed here to avoid redundancies.
It is worth to note that Eq. \eqref{eq:b_lgpc3} 
can also be expressed as $b^{\bullet}=K_1\big(3 a_2 h_1 h_3 -a_4\big)$ 
where $K_1$ represents the operator `$\tanh(\sin(\tanh(\cdot)))$'.
To shed light on the contribution of each term to $b^{\bullet}$, Fig.~\ref{fig:Viz_b_LGPC3} displays the temporal evolution of the actuation command
$b^{\bullet}$ and the relevant input from the states and from the harmonic functions.
It shows that the harmonic component $h_1 h_3$ destabilizes the stable oscillator by a quasi-periodic forcing 
while the states $a_2$ and $a_4$ act as an amplitude regulator.

\begin{figure}[htb]
	\centering
	\includegraphics[scale = 0.55]{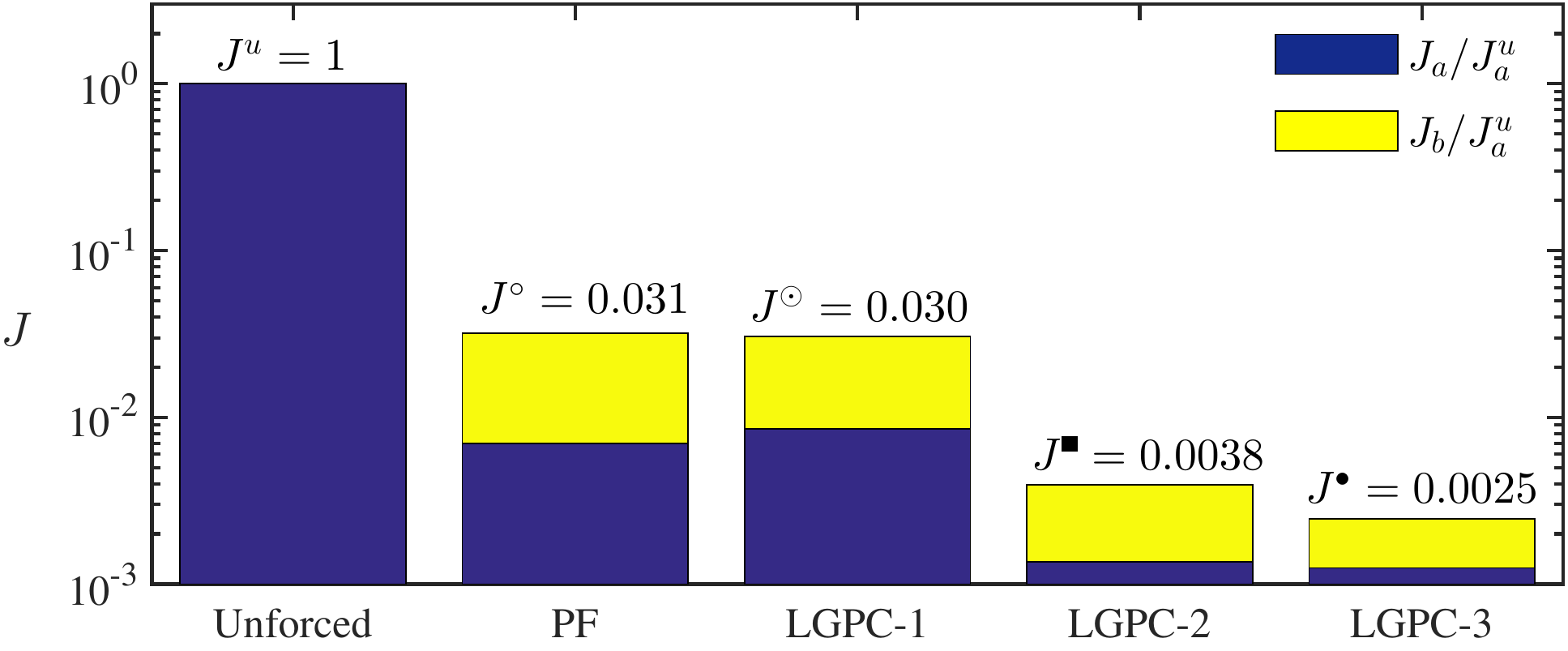}
	\caption{Synthesis of $J$ for different controls.}
	\label{fig:Bar_J_3osc}
\end{figure}

\begin{figure}[htb]
	\centering
	\includegraphics[scale = 0.7]{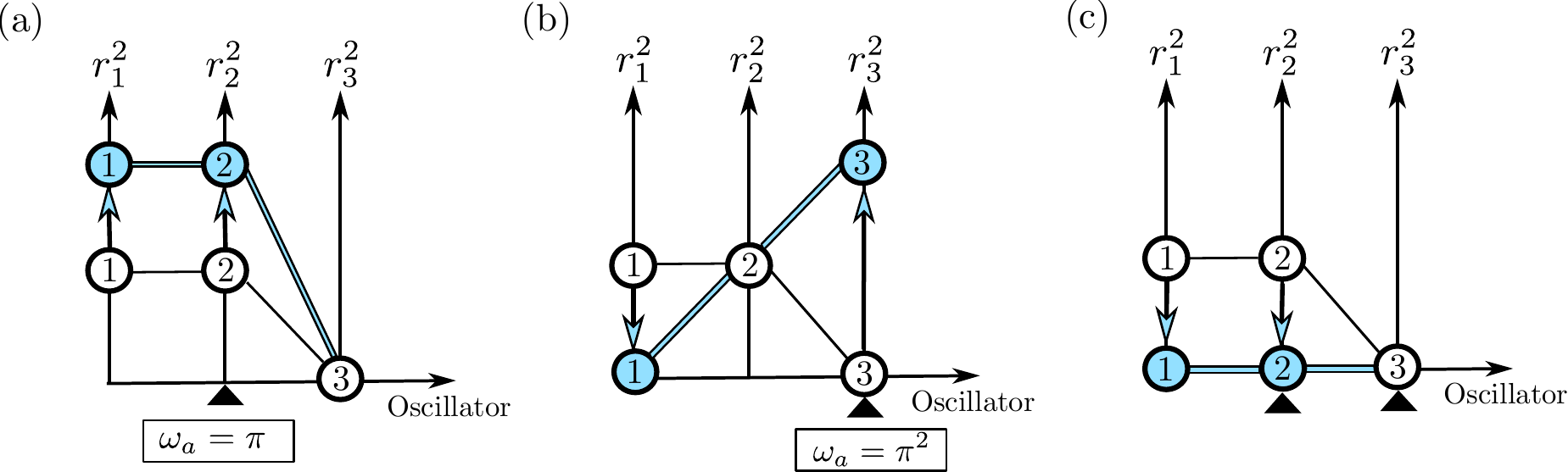}
	\caption{Synthesis of system dynamics under the forcing. The energy level of each oscillator is qualitatively indicated by circles. Unforced state: white circles connected by black line; forced state: colored circles connected by colored line. (a) Open-loop forcing at actuation frequency $\omega_a=\pi$. (b) Open-loop forcing at actuation frequency $\omega_a=\pi^2$. (c) Feedback control. The triangles indicate the oscillator(s) contributing to alter the first oscillator.
		The arrows show the transition state when control is applied.}
	\label{fig:sketch_energy}
\end{figure}

To summarize, optimal periodic forcing (PF), 
 open-loop multi-frequency forcing (LGPC-1),
full-state feedback (LGPC-2), 
and generalized feedback (LGPC-3) are compared.
The contributions to the cost function are depicted in Fig.~\ref{fig:Bar_J_3osc},
showing that the generalized feedback outperforms 
optimal periodic forcing and full-state feedback.
The stabilizing mechanisms are  schematically depicted in Fig.~\ref{fig:sketch_energy}.
\graphicspath{{./Figures/}}
\graphicspath{{./Figures/}}
\graphicspath{{./Figures/}}

\section{Drag reduction using LGPC}
\label{ToC:drag}
In this section, we apply LGPC to a turbulence control experiment targeting the drag reduction of a simplified car model.
Given that the drag of a ground vehicle is dominated by pressure drag, we aim to increase the base pressure and thus reduce the drag. 
For that, active control is applied on the wake flow using fluidic actuators.
In the following, the experimental setup is presented in Section \ref{ToC:Experimental setup}.
The implementation and results of LGPC are discussed in Section \ref{ToC:Ahmed_results_LGPC}.
Section \ref{ToC:Near wake} illustrates the effect of the optimal forcing on the near wake dynamics.

\subsection{Experimental setup}
\label{ToC:Experimental setup}
\begin{figure}[h]
	\centering
	\includegraphics[scale = 0.9]{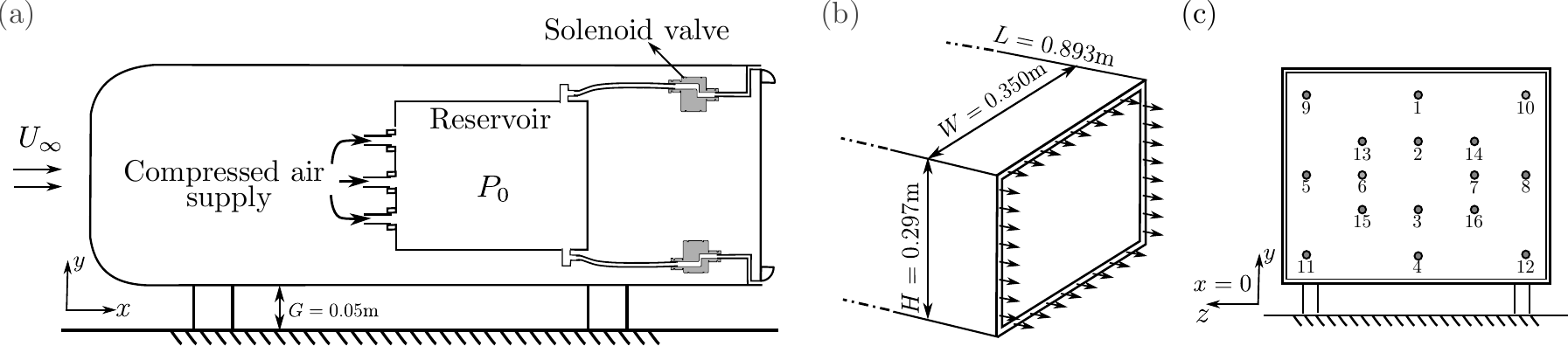}
	\caption{Experimental setup. (a) A slice of the model illustrating the actuation setup. (b) Side-view of pulsed jets. (c) Distribution and number of pressure sensors over the base surface. The first 12 pressure sensors are used for the feedback. $x,y,z$ represent the streamline, wall-normal and spanwise directions respectively. }
	\label{fig:setup}
\end{figure}

A sketch of the experimental setup is shown in Fig.~\ref{fig:setup}.
The experiment is performed in a closed-circuit wind tunnel, the test section of which is $\SI{2.4}{\meter}\times \SI{2.6}{\meter} \times \SI{6}{\meter}$.
The model is similar to the square-back Ahmed body \cite{Ahmed1984sae}
and 
has the following dimensions: height $H=\SI{0.297}{\meter}$, width $W=\SI{0.350}{\meter}$ and length $L=\SI{0.893}{\meter}$.
The ground clearance is set to $G=\SI{0.05}{\meter}\approx0.17H$ as in \cite{Ahmed1984sae}.
\graphicspath{{./Figures/}}
The experiment is conducted with a constant free-stream velocity $U_{\infty}=\SI{15}{\meter\per\second}$ corresponding to a Reynolds number $Re_{H}={U_\infty H}/{\nu}=3\times10^5$.
The wake is manipulated by pulsed jets emerging parallel to the free stream through the slits immediately beneath the trailing edges (see Fig.~\ref{fig:setup}(a) and (b)).
The slit thickness is $h_{\text{slit}}=\SI{1}{\milli\meter} \approx0.003H$.
In addition, a rounded surface of radius $9h_{\text{slit}}$ is installed immediately beneath each slit as an additional passive device.
The pulsed jets are driven by solenoid valves working in the frequency range $f\in[0,500]\si{\hertz}$, and are fed by a plenum connected to the lab pressurized air supply.
The actuation command $b$ is binary. The valves are closed at $b=0$ and open at $b=1$.
The flow is monitored by 16 pressure sensors distributed over the base surface, 12 of which are used as feedback sensors, see Fig.~\ref{fig:setup}(c).
Particle Image Velocimetry (PIV) is performed to capture the flow dynamics in the near wake and to identify the control effects.
The measured plane is the vertical (normal to ground) symmetry plane downstream the base.
The first and second order statistics of the streamwise (along $x$) and cross-stream (along $y$) velocity are computed based on 1000 images with a spatial resolution of 0.8\% of the model's height.
For more details on the experimental setup, see \cite{Barros2016jfm}.

\subsection{Results of LGPC}
\label{ToC:Ahmed_results_LGPC}
In the following, we apply LGPC on the plant for the purpose of increasing the base pressure.
We define the cost functional $J$ as
\begin{equation}
J=\frac{C_{p_b}^a}{C_{p_b}^u},
\label{eq:Jdrag}
\end{equation}
where $C_{p_b}^a$ and $C_{p_b}^u$ represent the time- and area-averaged base pressure coefficients in the actuated and unforced flow, respectively.
For estimating these quantities, all the pressure sensors in the base surface are used.
By definition, $J=1$ for the unforced flow.
$J<1$ ($J>1$) represents the increase (decrease) of the base pressure.

The optimal periodic forcing $b^\circ$ is found at $St^\circ_{H}=f^\circ H/U_\infty=6.6$ with duty cycle $DC^\circ=33\%$, resulting in $J=0.66$ and  increasing the base pressure by 33\%.
This result is taken as the benchmark.
The included sensors are $\vec{s^\prime}=[s_1^\prime,\ldots,s_{12}^\prime]$, where $s_i^\prime$ is the fluctuating component of the $i$th pressure sensor signal.
As the control command $b$ is binary, 
we apply a Heaviside function $\text{H}$ to transform the continuous output of a control law to a binary output, 
\textit{i.e.} \ $b=\text{H}(K(\vec{s^\prime}))$ where $\text{H}(x)=0, \text{if} ~x\leqslant0; \text{H}(x)=1$, otherwise. 
The control law is evaluated for a time period of $T=\SI{10}{\second}$. 
This value is approximately 500 convective time units defined by $H/U_\infty$.
This period has been found to be sufficient for good statistical accuracy \cite{Barros2015phd}.

First, we explore the open-loop multi-frequency control (LGPC-1)
optimizing the frequency combination.
Let $\vec{h}$ comprise 9 harmonic functions $h_i(t)=\sin(2\pi f_i t), i=1,\ldots,9$ listed in Table \ref{tab:h_i}.
In this case, the control law reads $b=\text{H}(K(\vec{h}))$.
Up to $N=4$ generations with $M=50$ individuals in each are evaluated.
We stop at the fourth generation because half of the individuals have similar $J$ values near the optimal one.
\begin{table}
	\caption{Harmonic functions $h_i(t)=\sin(2\pi f_i t)$ used as inputs of LGPC-1.}
	\begin{ruledtabular}
		\centering
		\begin{tabular}{l*{8}{c}r}
			Controller input  & $h_1$& $h_2$& $h_3$& $h_4$& $h_5$& $h_6$& $h_7$& $h_8$& $h_9$  \\ \hline
			$f_i$ (\si{\hertz})  & 10& 20& 50& 100& 200& 250& 333& 400& 500  \\
			$St_{H_i}$  & 0.2& 0.4& 1& 2& 4& 5& 6.6& 8& 10  \\ 
		\end{tabular}
		\label{tab:h_i}
	\end{ruledtabular}
\end{table}
The optimal control law reads:
\begin{equation}
b^{\odot}=\text{H}(h_5/h_8-0.622).
\label{eq:b_drag_LGPC2}
\end{equation}
The resulting cost $J^\odot=0.65$ beats the optimized periodic forcing, 
leading to 35\% base pressure recovery associated with 22\% drag reduction.
The actuation energy defined by the time-averaged momentum of pulsed-jets is about 7\% for both control laws.
The optimal control law contains two frequencies, indicating that LGPC-1 explores a multi-frequency forcing which outperforms the reference periodic forcing.

The results for LGPC-2, $b=\text{H}(K(\vec{s^\prime}))$, have been discussed in an earlier study \cite{Li2016ef2} and are not shown here.
Intriguingly, LGPC-2 provides a sensor optimization 
by reproducibly selecting only one sensor $s_4^\prime$ 
near the centre of bottom edge in the optimal control law.
The corresponding control emulates the optimal high-frequency periodic forcing but is slightly worse ($J^{\smallblacksquare}=0.72$).
A similar observation has been made for stabilization of the mixing layer \cite{parezanovic2015FTC},
where optimized high-frequency periodic forcing has outperformed GPC-optimized sensor-based feedback
in stabilizing the flow. 
At high frequencies, time delays and noise in sensor-based feedback
give rise to low-frequency actuation components which are detrimental to the cost function.
We could even change the dynamical system \eqref{eq:3osc}
to have an unbeatable periodic forcing, 
as discussed at the end of Section~\ref{ToC:3_osc_LGPC_2}.

Finally, a test of the generalized non-autonomous control LGPC-3 is performed
by combining the sensors $\vec{s^\prime}$ and 
the optimal harmonic forcing $h^\circ(t)=\sin(2\pi f^\circ t)$, i.e. $b=\text{H}(K(\vec{s^\prime},h^\circ))$.
LGPC-3 converges quickly to the optimal periodic forcing $b^\circ$.
The finding is in agreement with the LGPC-2 result 
where the optimal control emulates the optimal periodic forcing but is slightly worse.
LGPC-3 prefers to select the optimal periodic forcing to the sensor feedback. 
Upon these results, we do not pursue LGPC-3 $b=\text{H}(K(\vec{s^\prime},\vec{h}))$ by including multiple frequencies in this experiment.
We assume the result will be the same with LGPC-1.

In summary, 
LGPC identifies an open-loop multi-frequency forcing as the best control for drag reduction.
The underlying dynamics will be presented in the following section.
Note that this control has been identified by testing only 200 individuals in less than one hour.
The required optimization time is less than that for finding the best frequency and duty cycle for the periodic reference with a thorough parameter scan.

\subsection{Near wake dynamics of LGPC-1}
\label{ToC:Near wake}
\begin{figure}[h]
	\centering
	\includegraphics[scale = 0.54]{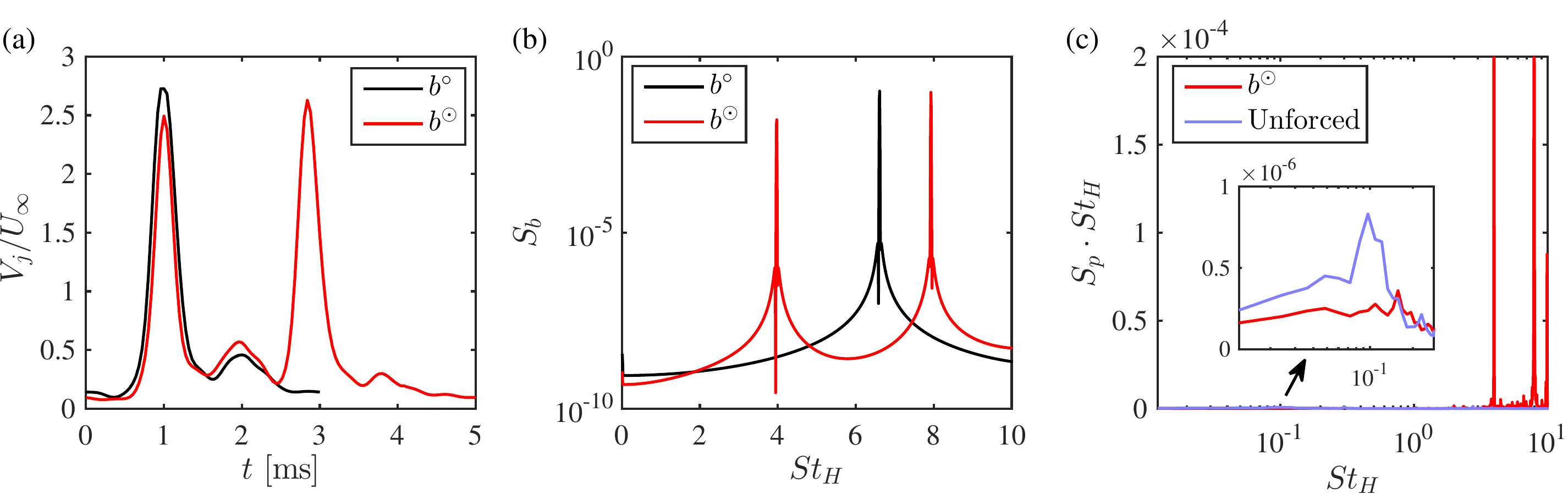}	
	\caption{ (a) Phase-averaged jet velocity $V_j$ for the optimal periodic forcing $b^{\circ}$ and the optimal LGPC-1 control $b^{\odot}$. (b) Power spectral density $S_b$ of $b^{\circ}$ and $b^{\odot}$. (c) Power spectral density $S_p$ of the area-averaged pressure coefficient.}
	\label{fig:spectra}
\end{figure}

In this section, we investigate the impact of the best control $b^{\odot}$ from LGPC-1 on the near wake dynamics.
To illustrate the actuation characteristics of $b^{\odot}$,
Fig.~\ref{fig:spectra} displays (a) its phase-averaged jet velocity over one period and (b) its power spectral density $S_b$.
The results of $b^\circ$ are also presented for comparison.
Intriguingly, $b^{\odot}$ exhibits a multi-frequency dynamic, 
showing two frequencies at $St_{H}=4$ and $St_{H}=8$, respectively.
\graphicspath{{./Figures/}}

It has been reported that forcing at frequencies several times that of the natural vortex shedding can stabilize the wake dynamics by inducing large dissipation and inhibiting the entrainment of fluid into the recirculation region \cite{Barros2016jfm,oxlade2015jfm}.
Here, LGPC-1 exploits similar actuations in an unsupervised manner.
The actuation frequencies in $b^{\odot}$ are one order of magnitude larger than that of the natural vortex shedding frequency $St_{H}^{\text{vs}}=0.2$.
The impact of the actuation on the wake dynamics can be further inferred from the base pressure fluctuation.
We use the area-averaged base pressure coefficient $\langle C_p \rangle$ as a global indicator of the dynamics.
Figure~\ref{fig:spectra}(c) compares the spectral energy of $\langle C_p \rangle$ for the unforced and optimal forced flow, where $S_{p}$ represents its power spectral density.
The high-frequency forcing has two major effects: (1) it significantly excites the frequencies over $St_{H}>1$, and (2) it suppresses a range of frequencies below $St_{H}<0.2$.
The high level of energy around $St_{H}=0.1$ in the unforced flow is associated with the bubble pumping frequency, which is induced by an axial oscillation of the recirculation bubble \cite{Berger1990JFS}.
It seems that the damping of this pumping mode contributes to reduce the drag.
The benefit in drag reduction by the suppression of this mode has been also observed in  \cite{khalighi2001sae}.
This result is a good illustration of the frequency crosstalk between low- and high-frequency, and corroborates the mechanisms proposed in \cite{oxlade2015jfm}.

\begin{figure}[h]
	\centering
	\includegraphics[scale = 0.9]{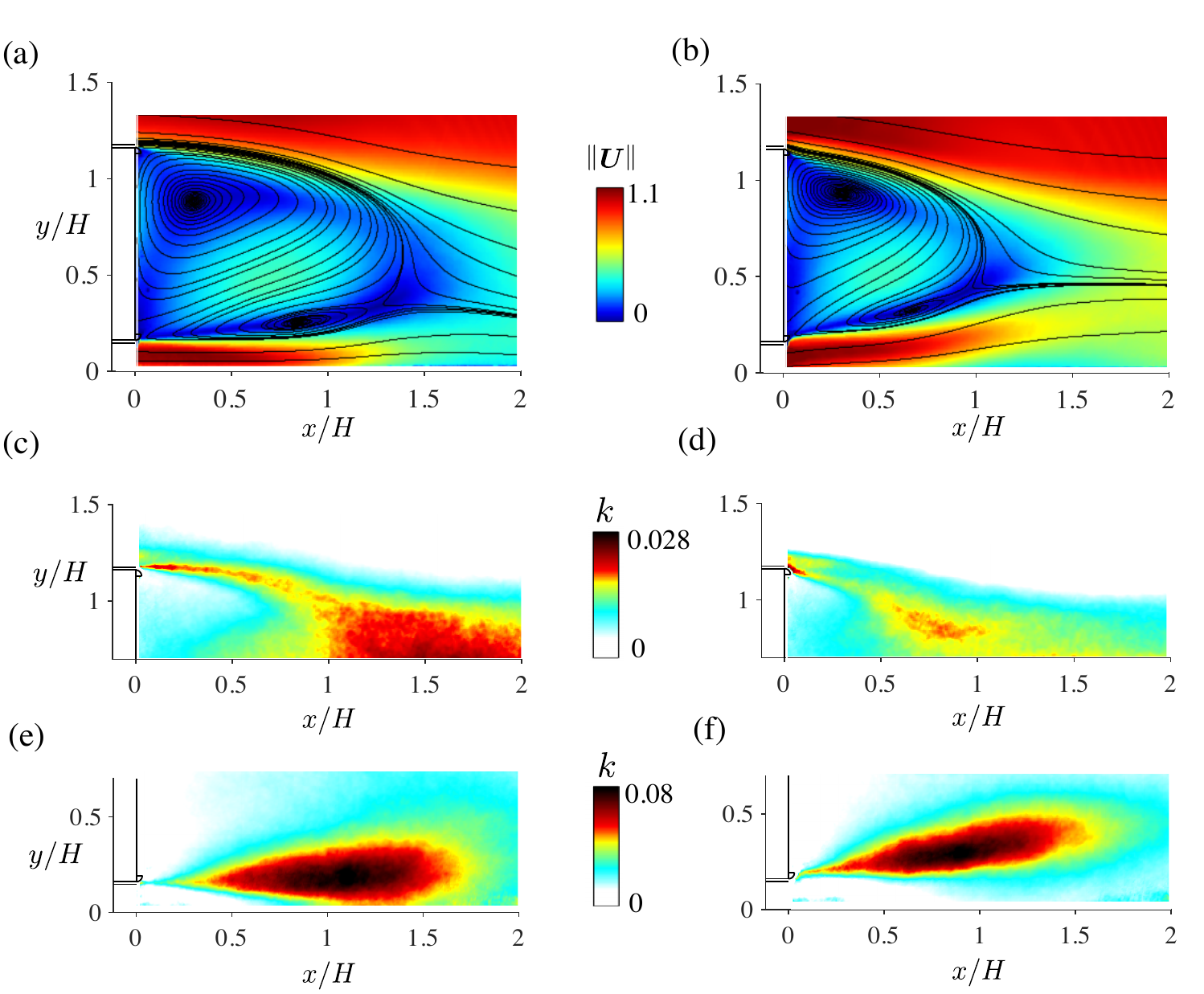}
	\caption{Near wake dynamics for the unforced baseline flow (a,c,e) and forced flow (b,d,f). (a,b) Time-averaged velocity norm $\norm{\vec{U}}$ and 2D streamlines; (c,d) 2D estimation of the turbulent kinetic energy $k$ for the upper shear layer; (e,f) $k$ for the lower shear layer.}
	\label{fig:PIV_U_k}
\end{figure}

Now, we focus on the effects of the best LGPC-1 control $b^\odot$ on the wake dynamics identified from the PIV measurements.
Figure~\ref{fig:PIV_U_k} shows the color map of the time-averaged velocity norm $\norm{\vec{U}}=\sqrt{\overline{u}^2+\overline{v}^2}$ overlapped with 2D streamlines (a, b) and 2D estimation of the turbulent kinetic energy $k=\frac{1}{2}(\overline{{u^\prime}^2}+\overline{{v^\prime}^2})$ (c-f) for the baseline (a,c,e) and controlled flow (b,d,f).
$\overline{u}$ and $\overline{v}$ represent the time-averaged streamwise and cross-stream velocity,  respectively.
$u^\prime$ and $v^\prime$ are their corresponding velocity fluctuations.
The values of these quantities are normalized by $U_{\infty}$.
\graphicspath{{./Figures/}}
The mean wake of the baseline flow consists of two counter-rotating structures with very low velocity inside, leading to a recirculating bubble extending up to $L_{r}/H\approx1.42$, where 
$L_r=\max_x(\overline{u}(x)=0)$
denotes the bubble length.
The upper recirculating structure dominates the wake and results in an asymmetry in the cross-stream direction.
The distribution of $k$ is concentrated in the shear layers, indicating its important role in the wake dynamics.
In addition, higher values of $k$ are noticeable at the lower shear layer near the ground which corroborates the asymmetry observed above.
Such asymmetry is ascribed to the presence of ground as a perturbation.

\begin{figure}[h]
	\centering
	\includegraphics[scale = 0.4]{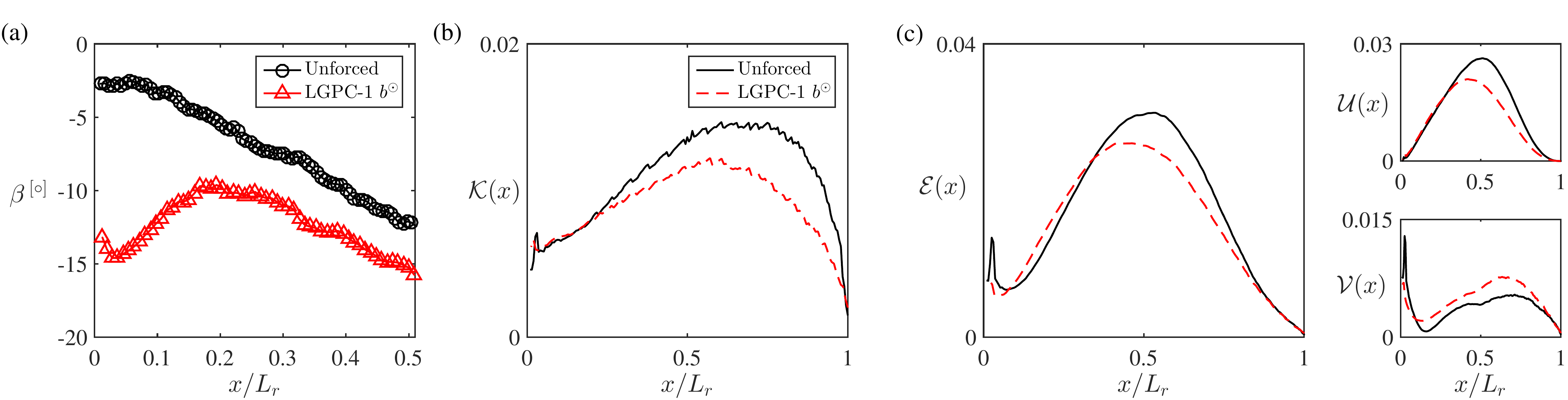}
	\caption{Effects of high-frequency forcing on the shear layer deviation and wake dynamics. (a) Evolution of the velocity angle $\beta$ along the streamline emerging from $(x/H,y/H)=(0.033,1.198)$; (b,c) streamwise evolution of $\mathcal{K}$ and $\mathcal{E}$. }
	\label{fig:Fig_angle_Tke_Ake}
\end{figure}

The forcing induces significant changes in the wake. 
First, the shear layers are highly deviated toward the model base, resulting in a thinner and shorter recirculation bubble, the length of which is $L_{r}/H\approx1.06$, reduced by 25\% compared with the baseline flow.
The vectorization of the shear layer is highlighted in Fig.~\ref{fig:Fig_angle_Tke_Ake}(a) by the velocity angle $\beta$ of the streamline emerging from the point $(x/H,y/H)=(0.033,1.198)$ located near the upper separating edge.
The angle variation immediately downstream the trailing edge ($x/H<0.1$) indicates that there is a reversal in the sign of streamline curvature.
This modification of curvature results in a local rise in base pressure.
Second, the vectorization of shear layers is accompanied by an overall reduction of turbulent kinetic energy inside the recirculation bubble, which can be qualitatively observed in Fig.~\ref{fig:PIV_U_k}(d) and (f).
Following the analyses in \cite{Barros2016jfm}, we quantify the modification of the wake dynamics by evaluating the streamwise evolution of the integral of the turbulent kinetic energy $\mathcal{K}(x)$ and averaged kinetic energy $\mathcal{E}(x)$ inside the domain $\Omega_{(\overline{u}<0)}$ defined as follows:
\begin{equation}
\mathcal{K}(x)=\int_{\Omega_{(\overline{u}<0)}}k(x,y)dy,
\label{eq:K_x}
\end{equation}
\begin{equation}
\mathcal{E}(x)=
\underbrace{\int_{\Omega_{(\overline{u}<0)}}\dfrac{\overline{u}^2(x,y)}{2}dy}_{\mathcal{U}(x)}
+
\underbrace{\int_{\Omega_{(\overline{u}<0)}}\dfrac{\overline{v }^2(x,y)}{2}dy}_{\mathcal{V}(x)}.
\label{eq:E_x}
\end{equation}
The results are shown in Fig.~\ref{fig:Fig_angle_Tke_Ake} (b) and (c).
We observe an overall reduction of $\mathcal{K}$ in the forced flow from $x/L_r=0.25$, indicating an attenuation of the fluctuating dynamics in the wake.
In particular, the significant reduction of $\mathcal{K}$ near the end of the mean recirculating bubble is believed to be linked with the very strong damping of the low frequency dynamics observed in Fig.~\ref{fig:spectra}(c).
A decrease of $\mathcal{E}$ is discernible very close to the base ($x/L_r<0.08$) and further downstream $x/L_r>0.33$. 
Between these two bounds, there is a slight increase of $\mathcal{E}$.
To gain insights into this evolution, we present separately the contribution of streamwise velocity $\mathcal{U}(x)$ and cross-stream velocity $\mathcal{V}(x)$ to $\mathcal{E}(x)$.
The decrease of $\mathcal{E}$ in the range $x/L_r<0.08$ is directly related to the reduction of $\mathcal{V}$ near the base, indicating that the upward flow adjacent to the base is less energetic in the forced flow.
Further downstream, $\mathcal{V}$ increases compared with the baseline flow.
In fact, the prominent deviation of the bubble boundary pushes the flow toward the inner wake and thus increases the absolute value of cross-stream velocity.
Correspondingly, we observe an increase of $\mathcal{E}$ in the range $x/L_r\in[0.08,0.33]$.
Beyond $x/L_r=0.33$, the decrease of $\mathcal{U}$ is amenable to  
the diminution of $\mathcal{E}$.
The overall attenuation of $\mathcal{U}$ indicates that the streamwise motion of the reversed flow is reduced by the forcing.

These observations show that a base pressure recovery is associated with: (1) the modification of streamline curvature which narrows and shortens the bubble and (2) the stabilization of the wake induced by the enhanced interaction of the small- and large-structures due to the high-frequency forcing.
These mechanisms are consistent with the results in \cite{Barros2016jfm} except that they did not observe a shorter bubble.
This difference is related to the actuation parameters.  
We actuate at a lower frequency and higher amplitude, yielding a higher angle deviation which is responsible for reducing the bubble length.

\graphicspath{{./Figures/}}

\section{Visualization of control laws}
\label{ToC:Visualization}

In this section,   
we illustrate the control laws and cost function values 
by an easily interpretable  'topological landscape',
generalizing earlier work \cite{Kaiser2017tcfd}.
First (Section \ref{ToC:Visualization:Method}),
 the visualisation technique is described,
employing a control-law distance metric and multidimensional scaling for feature extraction.
Then, (Section \ref{ToC:Visualization:Results}), the LGPC laws for the dynamical system
and the turbulence control experiment are depicted.

\subsection{Multidimensional scaling}
\label{ToC:Visualization:Method}
 
LGPC systematically explores the control law space by generating 
and evaluating a large number of control laws from one generation to the next.
An assessment of the similarity of control laws 
gives additional insights into their diversity and convergence to optimal control laws, 
\textit{i.e.}\ into the explorative and exploitative nature of LGPC.
For that purpose, we rely on Multidimensional Scaling (MDS) \cite{Mardia1979book},
a method classically used to visualize abstract data in a low-dimensional space.
The main purpose of MDS is to visualize the (dis)similarity of objects or observations. 
MDS comprises a collection of algorithms
to detect a meaningful low-dimensional embedding given a dissimilarity matrix.
Here, we employ Classical Multidimensional Scaling (CMDS) 
which originated from the works 
of \cite{Schoenberg} and \cite{YoungHouseholder}.

Let us define $N_K$ as the number of objects to visualize, 
and $\vec{D}=\left(D_{ij}\right)_{1\le i,j \le N_K}$ 
as a given distance matrix of the original high-dimensional data.
The aim of CMDS is to find a centred representation of points 
$\vec{\Gamma} = [\vec{\gamma}_1\quad \vec{\gamma}_2\quad \ldots\quad\vec{\gamma}_{N_K}]$
with $\vec{\gamma}_1,\ldots,\vec{\gamma}_{N_K}\in\mathbb{R}^r$,
where $r$ is typically chosen to be 2 or 3 for visualization purposes, 
such that the pairwise distances of the points approximate the true distances, 
\textit{i.e.}\ $\vert\vert \vec{\gamma}_i-\vec{\gamma}_j\vert\vert_2 \approx D_{ij}$.
The details of the implementation are given in Appendix \ref{Sec:AppA:CMDS}. 

We  choose  to visualize all control laws in a two-dimensional space $r=2$.
Thus, the number of objects is $N_K=M\times N$,
where $M$ is the number of individuals in a generation, and $N$ is the total number of generations. 
The distance between two control laws $b_i$ and $b_j$, $i,j \in \{1,\ldots,N_K\}$ 
shall measure their `effective difference'.
Let us consider the non-autonomous feedback $b_i = b_i \left (\vec{s}_i, \vec{h}_i \right)$.
Here, $\vec{s}_i(t)$ denotes the  sensor reading 
and  $\vec{h}_i(t)$ the harmonic control input 
on the corresponding $b_i$-forced attractor.
The squared difference between $b_i$ and $b_j$ is defined as
		\begin{equation}
		D_{ij}^2 =   \frac{
                \left \langle
		\left\vert b_i \left( \vec{s}_i(t),  \vec{h}_i(t)\right)
		- b_j \left( \vec{s}_i(t), \vec{h}_i(t)  \right)
		\right\vert^2 
                +
		\left\vert b_i \left( \vec{s}_j(t), \vec{h}_j(t)  \right)
		- b_j \left( \vec{s}_j(t),  \vec{h}_j(t)  \right)
		\right\vert^2 
		\right \rangle }{2}
		+ \alpha\,\vert J_i-J_j\vert.
		\label{Eq:DistanceMatrix1}
		\end{equation}
The time average  $\langle \cdot \rangle$ is taken over all sensor readings and corresponding harmonic input in the evaluation time interval 
from both forced attractors under control laws $b_i$ and $b_j$. 
Thus, $D_{ij}^2$  represents
the difference between the $i$th and $j$th control law
averaged over the sensor readings of both actuated dynamics.
The permutation of control laws $b_i$ and $b_j$ 
with its arguments
guarantees that the distance matrix is symmetric.
More importantly, this ensures that the control laws are compared 
in the relevant sensor space with an equal probability of both forced attractors.

The second term in \eqref{Eq:DistanceMatrix1} 
penalizes the difference of their achieved costs $J$ with coefficient $\alpha$.
The penalization coefficient $\alpha$ is chosen as the ratio 
between the maximum difference of two control laws 
(first term of $D_{ij}^2$)
and the maximum difference of the cost function 
(second term of $D_{ij}^2$).
Thus, the dissimilarities between control laws and between the cost functions 
have comparable weights in the distance matrix $D_{ij}$.
This penalization evidently smoothes the control landscape $J(\vec{\gamma})$.

A problem may arise for the comparison of two pure open-loop forcings $b_i$ and $b_j$.
We expect, for instance, that $b_i= \cos t$ and $b_j=\sin t$ give rise to the same actuation response 
modulo a time shift $\tau = \pi/2$ and 
would consider these control laws as equivalent.
Even for sensor-based feedback enriched by harmonic input, 
we expect the actuation response to be 'in phase' or synchronized with the harmonic input.
This expectation is taken into account  by 
minimizing the difference between two control commands modulo a minimizing time shift:
\begin{equation}
 D_{ij}^2 =   \mathop{\min}_{\tau} 
                \frac{
                \left \langle
		\left\vert 
                  b_i \big( \vec{s}_i(t), \vec{h}_i(t)\big)
		- b_j \big( \vec{s}_i(t-\tau), \vec{h}_i(t-\tau)\big)
		\right\vert^2
                +
		\left\vert 
                  b_i \big( \vec{s}_j(t), \vec{h}_j(t)\big)
		- b_j \big( \vec{s}_j(t-\tau), \vec{h}_j(t-\tau)\big)
		\right\vert^2
		\right \rangle }{2}
		+ \alpha\,\vert J_i-J_j\vert.
		\label{Eq:DistanceMatrix2}
\end{equation}
Evidently,  \eqref{Eq:DistanceMatrix1} and \eqref{Eq:DistanceMatrix2} concide at $\tau=0$.

Summarizing,
the square of the distance matrix $\boldsymbol{D}^2 = \left( D_{ij}^2 \right)$ is defined as follows:
\begin{itemize}
	\item[(1)]If both control laws have non-trivial harmonic input (are non-autonomous), 
                   \eqref{Eq:DistanceMatrix2} defines the distance.
	\item[(2)]Otherwise,
                   \eqref{Eq:DistanceMatrix1} is employed.
\end{itemize}
 
Applying CMDS to the distance matrix $\boldsymbol{D}$, each control law $b_i$ is associated with a point $\vec{\gamma}_i=(\gamma_{i,1},\gamma_{i,2})$ such that the distance between different $\vec{\gamma}_i$ emulates the distance between control laws defined by \eqref{Eq:DistanceMatrix1} and \eqref{Eq:DistanceMatrix2}.
More generally, $\vec{\gamma}_i$  are feature vectors which coefficients represent those features 
that contribute most on average to the discrimination of different control laws.

\subsection{Control landscapes for the LGPC runs}
\label{ToC:Visualization:Results}

\graphicspath{{./Figures/}}
\begin{figure}[h]
	\centering
	\includegraphics[scale = 0.45]{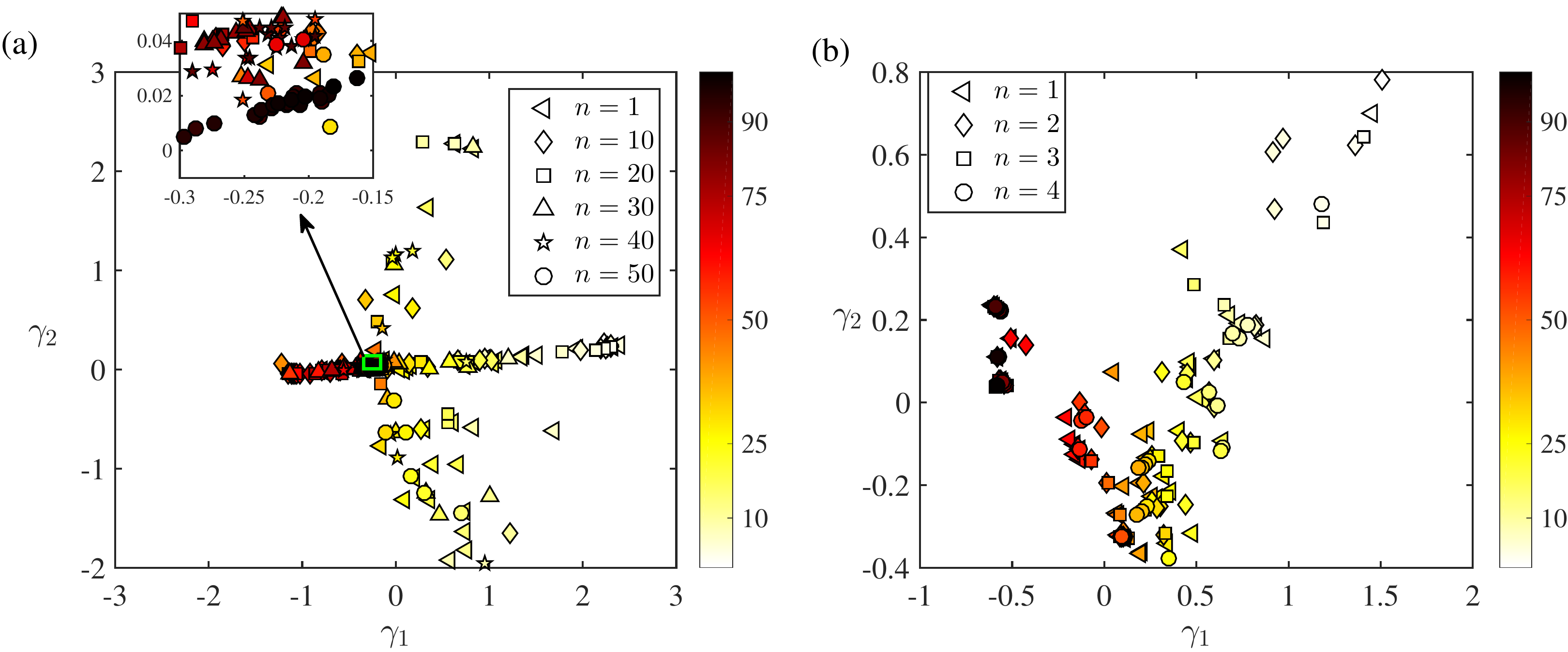}
	\caption{Visualization of the control laws obtained for (a) the three-oscillator model by LGPC-3 (Sec.~\protect{\ref{ToC:LGPC results}}) and (b) the simplified car model by LGPC-1 (Sec.~\protect{\ref{ToC:drag}}). $n$ represents the generation number. The color scheme corresponds to the percentile rank of the control laws with respect to their performance $J$. Darker color presents better performance. Control law $b_i$ is presented by the point $\vec{\gamma}=(\gamma_{1},\gamma_{2})$. The distance between two control laws, \textit{i.e.}\ two points, approximates their respective dissimilarity.}
	\label{fig:Viz_3_osc_LGPC2}
\end{figure} 

Figure~\ref{fig:Viz_3_osc_LGPC2} visualizes the control laws determined by LGPC-3 
for the three-oscillator model (a), and LGPC-1 for the simplified car model (b).
Due to the huge number of control laws in the three-oscillator model ($N_K=500\times50=25000$), we present every 10th individual in every 10th generation for clarity.
The full ensemble of individuals are shown for the simplified car model as its number is moderate ($N_K=50\times4=200$).
Each symbol represents a control law which is color-coded with respect to its performance ranking, 
for instance the dark color represents the best 10\% of the presented control laws.
The control laws in the first generation cover a significant portion of the control space, like in a Marte-Carlo search.
When the value of $n$ increases, we observe a global movement of control laws towards the minimum 
where better performance is obtained (darker color).
Moreover, the distances between control laws of different generations are also decreased resulting in a dense distribution.
This is illustrated in Fig.~\ref{fig:Viz_3_osc_LGPC2} (a) where the inserted figure gives a close view of the control laws near the origin point, where the best control law(s) are found at $[\gamma_1,\gamma_2]\approx[-0.18,0.02]$.
These observations show that LGPC has effectively explored the control space, 
evidenced by the extended  distribution of control laws.
In summary, the visualization provides not only a simple and revealing picture of the exploration and exploitation characteristics of the control approach, but also inspires further improvement of the methodology.

\section{Conclusion}
\label{ToC:Conclusion}
We have demonstrated that 
\emph{linear genetic programming control (LGPC)}
is a simple yet effective  model-free control strategy 
for strongly nonlinear dynamics with frequency crosstalk,
i.e.\ a very challenge of reduced-order modeling and model-based control design due to the difficulties in the mathematical modelling of the nonlinearities and limited knowledge of flow.
LGPC is shown to discover and exploit the most effective 
nonlinear open- and closed-loop control mechanisms
in dynamical systems and turbulence control experiments
in an automated unsupervised manner without any model or knowledge of the plant.

Three categories of LGPC are proposed in this work, 
an open-loop multi-frequency control $b=K(\vec{h})$, named LGPC-1,
an autonomous sensor-based feedback control $b=K(\vec{s})$, termed LGPC-2, 
and a generalized non-autonomous control $b=K(\vec{s},\vec{h})$ 
comprising the sensors $\vec{s}$ and time-periodic functions $\vec{h}$, called LGPC-3. 
All of them are successfully applied to the stabilization 
of a forced nonlinearly coupled three-oscillator model  (Section \ref{ToC:LGPC results}).
The obtained control laws stabilize the first unstable oscillator 
by exploiting two frequency crosstalk mechanisms: 
(1) the excitation of the third oscillator by a hard 'kick' for a quick transient and 
(2) the suppression of the second oscillator to sustain the low fluctuation level of the target dynamics.
Following the quick transient, 
the first and second oscillators enter into a quasi-stable state at nearly vanishing fluctuation levels, 
so the full-state feedback hardly needs to actuate and the control command starts to vanish.
The whole system is stabilized with only a small investment of the actuation energy at the very beginning of the control.
Thus, LGPC-exploited control laws show a performance 
over the optimal open-loop control as both a lower fluctuation level and a lower actuation energy are obtained.
The example and explored controller demonstrate the vital importance of frequency crosstalk for control design.

LGPC is applied in a turbulence control experiment 
targeting drag reduction of a car model (Section \ref{ToC:drag}).
It finds that multi-frequency forcing beats optimized periodic forcing by 22\% over 19\%,
the past benchmark for this square-back Ahmed body configuration.
This performance increase of 3\% pays for almost half of the invested actuation energy.
Perhaps surprisingly, the maximum actuation frequency 
is about 33 times that of the von K\'arm\'an vortex shedding.
This high-frequency forcing leads to a broadband suppression 
in very low frequencies of base pressure signals 
and a global attenuation of averaged and turbulent kinetic energy in the near wake, 
resulting in a more stabilized wake.
On the other hand, 
the mean wake geometry is modified such that the shear layers are deviated towards the center, 
resulting in a shorter, narrower, more stream-lined shaped bubble.
The drag reduction is ultimately achieved by the combined effect 
of the wake stabilization and the shear layer deviation
and can legitimately be called \emph{fluidic boat tailing}.

One of the many benefits of LGPC is that it explores automatically 
the control space with little or no knowledge of the system being controlled.
Moreover, the LGPC-3 ansatz for the control law can make the evolutionary algorithm 
choose between sensor-based feedback, multi-frequency forcing and combinations thereof.
In addition, 
the number of control laws evaluations for the Ahmed body drag reduction
was quite comparable to a single frequency optimization
but yields a much more general multi-frequency actuation
which is hardly accessible to a parametric study. 
In an even more general ansatz, 
noise signals $\vec{n}$  could also be included in the control law arguments, 
leading to $\vec{b}=\vec{\textbf{K}}(\vec{s},\vec{h},\vec{n})$.
Thus  stochastic forcing and its generalizations are included.
Another generalization is the use of temporal filters as considered operations. 
In \cite{Duriez2016book}, 
a filter-enriched GPC has successfully discovered the optimal linear quadratic gaussian control 
for the stabilization of a noise-driven oscillator.
In summary, LGPC can work on a search space 
which includes in principle any perceivable control logic 
with finite amount of operations.

Visualization of the ensemble of the control laws in a  two-dimensional plane sheds light 
on the explorative and exploitative nature of LGPC, 
and thus addresses the need to monitor the search space 
and guide the improvement of the algorithm.
The example given in Fig.~\ref{fig:Viz_3_osc_LGPC2} indicates 
clearly the search space topology and distills the local extrema in this feature space.
Evidently, in a future development of LGPC, 
this feature space may help to estimate the cost function of an untested control law
or be used to avoid the redundant testing of control laws in unpromising terrain. 
Thus, experimental testing time can be reduced.
The visualization is becoming an important component of LGPC 
for on-line decisions during a control experiment.

The authors currently improve the LGPC methodology, 
and pursue car model experiments 
for reducing the drag and yaw moment during cross-wind gusts.
LGPC opens refreshingly new paths in fluid mechanics, 
as estimation, prediction and control tasks 
are all regression problems miminizing a cost function.
LGPC exploits that control is a mapping from the plant 
sensors (output) to actuations (input)
optimizing aerodynamic or other goals.
Prediction is the mapping from the state to its time derivative or future state.
And estimation maps sensor signals to flow fields.
Evidently all these tasks can be solved with LGP.
Moreover, a single LGPC run yields already rich actuation response data 
for the computation of a control-oriented nonlinear black-box model.
Another more challenging direction is the exploitation of Navier-Stokes 
based insights in the problem formulation of LGPC.
LGPC and machine learning in general can reasonably be expected 
to be a game changer in future flow control and in fluid mechanics in general.

\begin{acknowledgments}
	We warmly thank the great support during the experiment by J.-M. Breux, J. Laumonier, P. Braud and R. Bellanger.
	The thesis of RL is supported by PSA Peugeot Citro\"{e}n in the context of OpenLab Fluidics (fluidics@poitiers).
	We also acknowledge the funding of the former Chair of Excellence 'Closed-loop control of turbulent shear layer flows using reduced-order models' (TUCOROM, ANR-10-CHEX-0015) supported by the French Agence Nationale de la Recherche (ANR).
	LC acknowledges the funding of the ONERA/Carnot project INTACOO (INnovaTive ACtuators and mOdels for flow cOntrol).
	EK gratefully acknowledges funding by the Moore/Sloan foundations, the Washington Research Foundation and the	eScience Institute.
	We appreciate valuable stimulating discussions with: Diogo Barros, Steven Brunton, Thomas Duriez and Andreas Spohn.
\end{acknowledgments}
\appendix
\section{Classical multidimensional scaling (CMDS)}
\label{Sec:AppA:CMDS}
Classical multidimensional scaling (CMDS) is employed to visualize the similarity of control laws (see Sec.~\ref{ToC:Visualization}).
CMDS aims to find a low-dimensional representation of points $\boldsymbol\gamma_i$, $i=1,\ldots,N_K$, 
such that the average error between the distances between points $\boldsymbol\gamma_i$ and the elements of a given distance matrix $\boldsymbol{D}$, 
here emulating the distances between the time series of different control laws, is minimal.

In order to find a unique solution to CMDS, 
we assume that $\boldsymbol{\Gamma}=[\boldsymbol{\gamma}_1\quad \boldsymbol{\gamma}_2\quad \ldots\quad\boldsymbol{\gamma}_{N_K}]$ with $\boldsymbol{\gamma}_1,\ldots,\boldsymbol{\gamma}_{N_K}\in\mathbb{R}^r$ is centered, 
\textit{i.e.}\ $\boldsymbol{\Gamma}$ is a mean-corrected matrix with $1/N_K\,\sum_{i=1}^{N_K}\, \boldsymbol{\gamma}_i = [0 \ldots 0]^T$.
Rather than directly finding $\boldsymbol{\Gamma}$, we search for the Gram matrix 
$\boldsymbol{B}=\boldsymbol{\Gamma}^T\boldsymbol{\Gamma}$ that is real, symmetric and positive semi-definite.
Since $\boldsymbol{\Gamma}$ is assumed to be centred, the Gram matrix is the Euclidean inner product, and we have 
$D_{ij}^2=\vert\vert \boldsymbol{\gamma}_i-\boldsymbol{\gamma}_j\vert\vert_2^2=B_{ii} + B_{jj} - 2\,B_{ij}$.
In the first step of the classical scaling algorithm, the matrix $\boldsymbol{D}_2$ of elements $\left(D_2\right)_{ij} = -\frac{1}{2}D_{ij}^2$ is constructed.
Then, we form the \mysingleq{doubly centred} matrix  $\boldsymbol{B} = \boldsymbol{C}\boldsymbol{D}_2\boldsymbol{C}$, where $\boldsymbol{C}=\boldsymbol{I}_{N_K}-N_K^{-1}\boldsymbol{J}_{N_K}$ with $\boldsymbol{I}_{N_K}$ the identity matrix of size $N_K$ and $\boldsymbol{J}_{N_K}$ an $N_K\times N_K$ matrix of ones. The term \mysingleq{doubly centred} refers to the subtraction of the row as well as the column mean.
Let the eigendecomposition of $\boldsymbol{B}$ be $\boldsymbol{B} = \boldsymbol{V}\boldsymbol{\Lambda}\boldsymbol{V}^T$ 
where $\boldsymbol{\Lambda}$ is a diagonal matrix with ordered eigenvalues
$\lambda_1\geq \lambda_2 \geq\ldots\geq\lambda_{N_K}\geq 0$
and $\boldsymbol{V}$ contains the eigenvectors as columns.
Then $\boldsymbol{\Gamma}$ can be recovered from
\begin{equation}
\boldsymbol{\Gamma} = \boldsymbol{\Lambda}^{\frac{1}{2}}\boldsymbol{V}^T.
\end{equation}
Having only the distance matrix, the resulting representation is only defined up 
to a translation, a rotation, and reflections of the axes.
If the distance matrix is computed using the Euclidean distance and all eigenvalues are non-negative,
$\boldsymbol{\Gamma}$ can be recovered.
If $r<N_K$, there exist $N_K-r$ zero eigenvalues, 
in which case a low-dimensional subspace can be found where the presentation of $\boldsymbol{\Gamma}$ would be exact.
For other distance metrics, the distances of the presentation found by CMDS is an approximation to the true distances.
Some eigenvalues may be negative and only the positive eigenvalues and their associated eigenvectors 
are considered to determine an approximative representation of $\boldsymbol{\Gamma}$.
Note that for the Euclidean distance metric, CMDS is closely related to a principal component analysis (PCA)
commonly used to find a low-dimensional subspace. 
While CMDS, and multi-dimensional scaling generally, uses a distance matrix as input, PCA is based on a data matrix.
A distance matrix $\boldsymbol{D}$ can be directly computed for the centred matrix $\boldsymbol{\Gamma}$. 
If the Euclidean distance is employed for computing the distances, 
the result from applying CMDS to $\boldsymbol{D}$ corresponds to the result 
from applying PCA to $\boldsymbol{\Gamma}$.
A proof can be found in \cite{Mardia1979book}.
The quality of the representation is typically measured by 
$\sum_{i=1}^r\,\lambda_i/\sum_{i=1}^{N_K-1}\,\lambda_i$, 
and more generally if $\boldsymbol{B}$ is not positive semi-definite using
$\sum_{i=1}^r\,\lambda_i/\sum_{\lambda>0}\,\lambda_i$.

\bibliography{Main}

\end{document}